\documentstyle[12pt]{article}

\newcommand{\nn}{\nonumber}
\newcommand{\be}{\begin{equation}}
\newcommand{\ee}{\end{equation}}
\newcommand{\bea}{\begin{eqnarray}}
\newcommand{\eea}{\end{eqnarray}}
\renewcommand{\d}{\mathrm{d}}
\def\a{\alpha}

\def\d{\delta}

\def\s{\sigma}
\def\sb{\bar{\sigma}}

\def\k{\kappa}
\def\l{\lambda}
\def\m{\mu}
\def\n{\nu}

\def\p{\phi}
\def\pb{\bar\psi}

\def\th{\theta}
\def\tb{\bar{\theta}}

\def\s{\sigma}

\def\pt{\partial}

\def\ad{\dot\alpha}

\def\Db{\bar{D}}
\def\As{A^\ast}
\def\Fs{F^\ast}
\def\P{\Phi}
\def\Pd{\Phi^\dag}

\begin{document}

\title{Supersymmetric Galileons}

\author{
Justin Khoury${}^{1}$,
Jean-Luc Lehners${}^{2}$,
Burt A. Ovrut${}^{1}$
}
\date{}

\maketitle
\begin{center} {\small ${}^1${\it Department of Physics, University of
      Pennsylvania, \\ Philadelphia, PA 19104-6395, U.S.A.}\\[0.2cm]
      ${}^2${\it Max-Planck-Institute for Gravitational Physics (Albert-Einstein-Institute), \\
      D-14476 Potsdam/Golm, Germany }}\\

\end{center}

\abstract{Galileon theories are of considerable interest since they allow for stable violations of the null energy condition. Since such violations could have occurred during a high-energy regime in the history of our universe, we are motivated to study supersymmetric extensions of these theories. This is carried out in this paper, where we construct generic classes of ${\cal{N}}=1$ supersymmetric Galileon Lagrangians. They are shown to admit non-equivalent stress-energy tensors and, hence, vacua manifesting differing conditions for violating the null energy condition. The temporal and spatial fluctuations of all component fields of the supermultiplet are analyzed and shown to be stable on a large number of such backgrounds. In the process, we uncover a surprising connection between conformal Galileon and ghost condensate theories, allowing for a deeper understanding of both types of theories.}

\section{Introduction and Overview}

Matter in the universe is typically assumed to satisfy the Null Energy Condition (NEC)~\cite{Hawking:1973uf}. This is because standard two-derivative theories generically lead to the appearance of ghosts or gradient instabilities on NEC-violating backgrounds~\cite{Dubovsky:2005xd}. Moreover, higher-derivative theories --- generically associated with equations of motion that are of third- and higher-order in derivatives --- also lead to the appearance of ghosts and are, therefore, catastrophically unstable~\cite{Ostro}$-$\cite{Khoury:2006fg}.

However, in recent years it has become clear that these theoretical limitations are neither necessary nor, perhaps, desirable. Indeed, there are a number of cosmological situations in which violations of the NEC become inevitable, including inflationary models with extra dimensions~\cite{Wesley:2008fg,Steinhardt:2008nk}, string gas cosmological scenarios~\cite{Nayeri:2005ck}$-$\cite{Brandenberger:2006pr} (see~\cite{Battefeld:2005av} for a review), and pre-big bang~\cite{Gasperini:1992em} and ekpyrotic theories with cosmic bounces~\cite{Khoury:2001wf}$-$\cite{Lehners:2010ug} in which the universe reverts from contraction to expansion (see~\cite{Lehners:2008vx,Lehners:2010fy} for reviews of ekpyrotic/cyclic theories). Furthermore, there is an important caveat to the theorem of~\cite{Dubovsky:2005xd} --- namely, the existence of stable, NEC-violating higher-derivative theories that nevertheless lead to equations of motion with at most two derivatives acting on any field. Two classes of such theories have been studied in the literature so far. The first is ghost condensate theories~\cite{ArkaniHamed:2003uy}, in which the Lagrangian is taken to be an analytic function of a scalar field $\phi$ and $X\equiv -\frac{1}{2}(\pt\p)^2.$ The second example is provided by the Galileons~\cite{Deffayet:2001uk}$-$\cite{Andrews:2010km}, in which higher-derivative terms are combined precisely so that the equations of motion have at most two derivatives acting on each field. Both types of theories have the remarkable property that they allow for stable violations of the NEC and, hence, both types of theories can be used as effective theories to model novel cosmological scenarios. Indeed, ghost condensates have been used to violate the NEC~\cite{Creminelli:2006xe} and, hence, enable cosmic bounce from contraction to expansion in the New Ekpyrotic Scenarios in~\cite{Buchbinder:2007ad,Buchbinder:2007tw,Buchbinder:2007at}. Similarly, galileons have been used to devise cosmological scenarios in which the universe expands from asymptotically flat initial conditions~\cite{Nicolis:2009qm,Creminelli:2010ba}.

The relationship of these theories to string theory is not yet entirely clear, although it is interesting that (in a certain small field limit) the Galileon theories describe the fluctuations of a brane embedded in a higher-dimensional space-time~\cite{deRham:2010eu} (see also \cite{VanAcoleyen:2011mj} for a derivation of the Galileons via compactification of Lovelock gravity). Such scenarios arise naturally in heterotic M-theory~\cite{Lukas:1997fg}$-$\cite{Lukas:1998tt}, for example, where five-branes wrapped on holomorphic two-cycles~\cite{Lukas:1998hk}$-$\cite{Lima:2001nh} can exist in the five-dimensional bulk space. The visible sector of such theories can contain exactly the supersymmetric standard model~\cite{Braun:2004xv}$-$\cite{Anderson:2009mh} and, hence, present realistic vacua to explore Galileon cosmology. If ghost condensate or Galileon theories turn out to be relevant in modeling the dynamics of the universe in the high-energy regime, then it would seem necessary to consider these theories in a supersymmetric context. Quite independently, it is of theoretical interest to have a model that allows one to study the interplay between supersymmetry and NEC violation. In previous work~\cite{Khoury:2010gb}, we supersymmetrized the ghost condensate models. In this paper, we further this viewpoint by constructing generic ${\cal{N}}=1$ supersymmetric extensions of Galileon theories.

In~\cite{Khoury:2010gb}, we studied the ${\cal N}=1$ supersymmetric extension of ghost condensate theories using chiral multiplets. As reviewed in Sec.~\ref{GhostCond}, the extra scalar and auxiliary fields required by supersymmetry are well-behaved for such models, while the fermionic member of the supermultiplet is not. Specifically, the fermion kinetic term on the ghost condensate background violates Lorentz-covariance --- the spatial-derivative part has the wrong sign while the time-derivative part has the correct one. In this paper, we show that a manifestly supersymmetric interaction can be added to this theory which has the property of restoring the fermion kinetic term to its canonical form. This will be the subject of Sec.~\ref{fermioncure}. When we examine the effect of adding this term on the bosonic part of the theory, we find a surprise: the resulting scalar field theory is precisely the second- and third-order conformal Galileon theory! Hence, in rendering the fluctuations around the ghost condensate background canonical, we re-discover the second- and third-order conformal Galileon model. Moreover, using a field redefinition, the ghost condensate background is easily seen to be equivalent to the ``self-accelerating'' de Sitter solution of the Galileon theory. The real difference between these two theories lies in the form of the spatial gradient terms. Generally speaking, the Galileon theories are much better behaved with regard to spatial gradients than their ghost condensate counterparts, as will be discussed in Sec.~\ref{galP(X)}.

When we consider the fourth-order conformal Galileon Lagrangian, we find that there are now many choices in how to construct a supersymmetric generalization. Since these choices become vastly more numerous for the fifth- (and highest) order Galileon theory, we only present  terms up to fourth-order in this paper. There are two reasons for this proliferation. The first is that, using integration by parts, one may rewrite a given action into one that is equivalent up to total derivatives. If we now discard the total derivatives and supersymmetrize the new action, we generically end up with inequivalent results. This is an important ambiguity regarding higher-derivative theories --- that is, different theories, though related by integrations by parts (on a flat background), lead to different stress-energy tensors and, hence, different conditions for NEC violation. We provide a detailed treatment of these issues in Sec.~\ref{NECviolate}. The second reason for this proliferation of choices is intrinsically supersymmetric. As in our previous paper on ghost condensate theories, we construct supersymmetric extensions of many-field terms by using a number of smaller building blocks. And for some terms, there are several inequivalent ways of subdividing them into separate building blocks. It is interesting to note that the resulting supersymmetric extensions can be quite different, and can now contain non-canonical fluctuations of the fermion field for example. This will be discussed in detail in Sec.~\ref{susygal}.

In Sec.~\ref{discussion}, we discuss our results and speculate on future applications. Two appendices are included. The first provides useful formulae regarding the supersymmetric building blocks that we are using, and the second discusses in more detail those supersymmetric extensions that contain non-covariant kinetic terms for the fermionic fields.

\section{Supersymmetric Ghost Condensate} \label{GhostCond}

\subsection{A Review of Ghost Condensation}

The simplest form of a ``ghost condensate''~\cite{ArkaniHamed:2003uy} arises within the context of a single real scalar field $\phi$ in four dimensions. Assuming space-time is flat and non-dynamical, the  evolution of $\phi$ is governed by a higher-derivative Lagrangian of the form
\begin{equation}
{\cal{L}}=P(X) \ ,
\label{1}
\end{equation}
where $P(X)$ is an arbitrary function that is analytic around zero in
\begin{equation}
X \equiv -\frac{1}{2m^4}(\partial \phi)^2 = \frac{1}{2m^4}(\dot{\phi}^2-\phi^{,i} \phi_{,i}) \ .
\label{2}
\end{equation}
The mass scale $m$ is introduced to render $X$ dimensionless. To simplify notation, we set $m=1$ in most of the paper.
For purely time-dependent solutions, the associated equation of motion is given by
\begin{equation}
\frac{\rm d}{{\rm d}t}\left(P_{,X} \dot{\phi} \right)=0 \ .
\label{3}
\end{equation}
Clearly, $\phi={\rm const.}$ is a solution. However, (\ref{3}) also allows for solutions with arbitrary constant $X$, that is,
\begin{equation}
\phi=c\,t\,,
\label{4}
\end{equation}
where $c$ is a constant. Although in this paper space-time is taken to be non-dynamical, we note that in a cosmological context the equation of motion on a Friedmann-Robertson-Walker background becomes
\begin{equation}
\frac{\rm d}{{\rm d}t}\left(a^{3}P_{,X} \dot{\phi} \right)=0 \ ,
\label{5}
\end{equation}
where $a(t)$ is the scale factor of the universe. For a generic choice of $P(X)$, this implies that $\dot{\phi}$ must redshift as the universe expands.
However, there is one key exception: if $P(X)$ has an {\it extremum} at some $X = c^2/2$, then $\phi = c\,t$ is a solution to~(\ref{5}) independent of the behavior of $a(t)$. Moreover,
this solution is an attractor on an expanding background --- small departures away from the extremum are driven to zero by Hubble friction. This solution,
$\phi = c\,t$, spontaneously breaks Lorentz invariance and is called a ghost condensate.

Returning to a flat background and expanding in fluctuations
\begin{equation}
\phi=c\,t+\delta \phi(t,{\vec{x}})
\label{6}
\end{equation}
around a ghost condensate, to quadratic order in $\delta\phi$ the Lagrangian becomes
\begin{equation}
{\cal L}_{\rm quad} =
XP_{,XX}\cdot(\delta {\dot{\phi}})^2 - 0 \cdot \delta \phi^{,i}\delta\phi_{,i} \ .
\label{7}
\end{equation}
As a result of Lorentz-breaking, the coefficients in front of the time and spatial derivative terms are unequal. We see that the condition for the absence of a ghost is
\begin{equation}
X P_{,XX} >0\,,
\label{8}
\end{equation}
which is automatically satisfied close to a local {\it minimum} of $P(X)$.
(For a general $X = {\rm const.}$ solution, the ghost-free condition is $XP_{,XX} + P_{,X}/2 > 0$~\cite{Garriga:1999vw}.)
We, henceforth, assume this is the case. However, the vanishing of the second term in (\ref{7}) is troubling, since it clearly signals that the ghost condensate is on the verge of a gradient instability. Can this potential instability be removed? Happily, the answer is affirmative, although it requires introducing higher-derivative terms, such as~\cite{ArkaniHamed:2003uy}
\begin{equation}
-\frac{(\Box \phi )^{2}}{M^2}\ ,
\label{9}
\end{equation}
into the Lagrangian that are not of the $P(X)$ type. Such corrections are expected from an effective field theory point of view. Because this term involves two derivatives per field,
the background $\phi = c\,t$ clearly remains a solution. However, (\ref{9}) does affect the gradient term of the fluctuations, giving rise to the dispersion relation $\omega^2 \sim k^4/M^2$.
For large enough mass $M$, this higher-derivative term
can be consistently treated
as a small correction.
Be this as it may, the question of temporal ghosts and/or gradient instabilities in ghost condensate theories is an
important one, and will become even more important in the supersymmetric context.

Before proceeding, we introduce the following simplification. Sufficiently close to the ghost condensate point, $P(X)$ is approximately quadratic. Without loss of generality,
one can rescale the field $\phi$ so that the minimum lies at $X= 1/2$ (corresponding to $c=1$) and write the prototypical ghost condensate action as
\begin{equation}
{\cal{L}}=-X+X^2 =
+\frac{1}{2}(\partial \phi)^2 + \frac{1}{4}(\partial \phi)^4 \,.
\label{10}
\end{equation}
The quadratic Lagrangian (\ref{7}) now becomes
\begin{equation}
{\cal L}_{\rm quad} = (\delta {\dot{\phi}})^2 - 0 \cdot \delta \phi^{,i}\delta\phi_{,i} \ .
\label{11}
\end{equation}
We will use Lagrangian (\ref{10}), which contains all of the essential physics,  to supersymmetrize ghost condensate theories.

\subsection{Supersymmetric Ghost Condensate}

In \cite{Khoury:2010gb}, we presented an ${\cal{N}}=1$ supersymmetric extension of the bosonic ghost condensate theory in (\ref{10}).  To do this, consider a {\it chiral} superfield
\begin{equation}
\Phi = A + i\theta\sigma^\mu{\bar{\theta}} A_{,\mu} + \frac{1}{4}\theta\theta{\bar{\theta}}{\bar{\theta}} \Box A + \theta\theta F + \sqrt{2} \theta\psi -\frac{i}{\sqrt{2}}\theta\theta \psi_{,\mu}\sigma^\mu{\bar{\theta}} \,,
\label{12}
\end{equation}
with the complex scalar $A(x),$ the auxiliary field $F(x)$ and the spinor $\psi_\alpha (x)$ being functions of the ordinary space-time coordinates $x^\mu.$ Spinor indices which we do not write out explicitly are understood to be summed according to the convention $\psi \theta = \psi^\alpha \theta_\alpha$ and ${\bar{\psi}}{\bar{\theta}} = {\bar{\psi}}_{\dot{\alpha}} {\bar{\theta}}^{\bar{\alpha}}.$
The complex scalar is chosen so that
\begin{equation}
A=\frac{1}{\sqrt{2}}(\phi+i\chi) \ ,
\label{13}
\end{equation}
where $\phi$ is the real field of the bosonic condensate theory.
The imaginary component $\chi$ is a new real scalar degree of freedom, introduced into the condensate theory by supersymmetry. That is, $\phi$ is taken to be the lowest component of the ${\cal{N}}=1$ chiral supermultiplet $(\phi, \chi, \psi, F)$.

It was shown in \cite{Khoury:2010gb} that a supersymmetric extension of the prototypical ghost condensate Lagrangian (\ref{10}) is given by \footnote{As discussed in \cite{Khoury:2010gb}, there exists a second (inequivalent) supersymmetric extension of $X^2$, which however leads to the exact same issues with the fermionic fluctuations as those discussed below. Our discussion, and the cure proposed below, are thus general.}
 \begin{equation}
 {\cal L}^{\rm SUSY}=
 \left(-{\Phi}{\Phi}^{\dagger} +\frac{1}{16} D{\Phi}
 D{\Phi} {\bar{D}{\Phi}^{\dagger} {\bar{D}} {\Phi}^{\dagger}} \right)
\bigg\vert_{\theta \theta {\bar{\theta}} {\bar{\theta}}} \ ,
\label{14}
\end{equation}
where $|_{\theta \theta {\bar{\theta}} {\bar{\theta}}}$ indicates taking the ${\theta \theta {\bar{\theta}} {\bar{\theta}}}$-component of a superfield. (Here and throughout the paper, derivatives are understood as acting only on the nearest superfield, unless noted otherwise. For example, $D{\Phi} D{\Phi} {\bar{D}{\Phi}^{\dagger} {\bar{D}} {\Phi}^{\dagger}} = (D{\Phi})
 (D{\Phi}) ({\bar{D}{\Phi}^{\dagger})({\bar{D}} {\Phi}^{\dagger}})$. Similarly for space-time derivatives acting on component fields.)
In terms of component fields, (\ref{14}) becomes
\begin{eqnarray}
 {\cal{L}}^{\rm SUSY} &=& \frac{1}{2}(\partial\phi)^2 + \frac{1}{4}(\partial\phi)^4 +\frac{1}{2} (\partial\chi)^2 - \frac{1}{2}(\partial\phi)^2(\partial\chi)^2 + (\partial\phi \cdot \partial\chi)^2  \nn \\
 & - &\frac{i}{2}(\psi_{,\mu}\sigma^\mu {\bar{\psi}} - \psi \sigma^\mu {\bar{\psi}}_{,\mu}) -\frac{i}{4}(\partial\phi)^2(\psi_{,\mu}\sigma^\mu {\bar{\psi}} - \psi \sigma^\mu {\bar{\psi}}_{,\mu})
   \nn \\
  &-& \phi_{\mu}\phi_{,\nu}\frac{i}{2}(\psi^{,\nu}\sigma^\mu {\bar{\psi}} - \psi \sigma^\mu {\bar{\psi}}^{,\nu}) + \ldots
\label{15}
\end{eqnarray}
where we display terms to quadratic order only in $\chi$, $\psi$ and set $F=0$. Note that for $\chi=\psi=0$, this expression exactly reduces to Lagrangian (\ref{10}). It is in this sense that $ {\cal{L}}^{\rm SUSY}$ is the supersymmetric extension of the prototype bosonic condensate theory. Since $\chi$ always appears at least to quadratic order, it can consistently be set to zero. Thus the equations of motion can be solved by the ghost condensate
\begin{equation}
\phi=c\,t~, \quad \chi=0 \ .
\label{16}
\end{equation}
The classical fermion solution is, of course, zero. That is, the Lorentz-violating ghost condensate continues to exist as a vacuum of the supersymmetrized theory.

Setting $c=1$ and expanding in fluctuations
\begin{equation}
\phi=t+\delta \phi(t,{\vec{x}}) \,, \quad \chi=\delta \chi(t,{\vec{x}}) \,, \quad \psi=\delta \psi(t,{\vec{x}})
\label{17}
\end{equation}
around this vacuum, we find to quadratic order that
\begin{eqnarray}
{\cal L}^{\rm SUSY}_{\rm quad} &=& (\dot{\d\p})^2 - \, 0 \cdot \d\p^{,i}\d\p_{,i} \nn \\
& +& \, 0 \cdot (\dot{\d\chi})^2 + \d\chi^{,i}\d\chi_{,i}\nn \\
& +& \frac{i}{4}\left(\d\psi_{,0}\s^0 \d\bar\psi - \d\psi \s^0 \d\bar\psi_{,0}\right)  -\frac{i}{4}\left(\d\psi_{,i}\s^i \d\bar\psi - \d\psi \s^i \d\bar\psi_{,i}\right) \ .
 \label{20}
 \end{eqnarray}
The first line reproduces the standard result (\ref{11}) for the single $\phi$ field ghost condensate, as it must. That is, the time derivative term is ghost-free but, at the minimum of $P(X),$ the spatial gradient term for $\d\p$ vanishes. As discussed above, higher-derivative terms of the form (\ref{9}) cure the potential gradient instability in the bosonic theory and stabilize the dispersion relation. Can one find a supersymmetric generalization of these terms? In \cite{Khoury:2010gb} we showed that this can indeed be done. The simplest such example is
\be
-\frac{1}{2^{11}}D\P D \P \Db \Pd \Db \Pd \left(\{D,\Db\}\{D,\Db\}(\P + \Pd)\right)^2 \bigg\vert_{\th\th\tb\tb,\;{\rm quad}} =-(\Box \d\p)^2\,,
\label{21}
\ee
where we have evaluated this up to quadratic order in fluctuations around a ghost condensate background. To this order, (\ref{21}) does not contain $\chi$, $\psi$ or the auxiliary field $F$ at all.

Now consider the second line in ${\cal L}^{\rm SUSY}_{\rm quad}$. This is the kinetic term for the scalar fluctuation $\delta \chi$ and, hence, is new to the supersymmetric theory. Note that this suffers from {\it two} serious problems. The first is that the temporal derivative term vanishes and, hence, this field is marginally a ghost. Secondly, the spatial gradient term has the wrong sign. Fortunately, it was shown in \cite{Khoury:2010gb} that supersymmetric terms can be added to (\ref{14}) that solve both problems. These are, for example,
\bea
&& \bigg[\,\frac{8}{16^2}D \P D \P \Db \Pd \Db \Pd \left(\{D,\Db\}(\P - \Pd)\{D,\Db\}(\Pd - \P)\right) \nn
\\ &&- \frac{4}{16^3}D \P D \P \Db \Pd \Db \Pd \left(\{D,\Db\}(\P + \Pd)\{D,\Db\}(\P - \Pd)\right)^2 \bigg]\bigg\vert_{\th\th\tb\tb, \, {\rm quad}} \nn \\
&&=-2(\pt\p)^4(\pt\chi)^2 - (\pt\p)^4(\pt\p\cdot\pt\chi)^2 \,.
 \label{22}
\eea
Adding these to  Lagrangian (\ref{14}), and expanding to quadratic order around the ghost condensate, changes both the time and spatial gradients of $\chi$ in (\ref{20}) to the Lorentz-covariant expression
\be
{\cal L}^{\rm SUSY}_{\rm quad}=\dots  +({\d\dot\chi})^2 -\d\chi^{,i}\d\chi_{,i} +\dots
\label{24}
\ee
This renders the $\chi$ fluctuations stable, without adversely affecting anything else.
In particular, since (\ref{22}) vanishes when $\chi$ is set to zero, the sum of (\ref{14}) and the superfield expression in (\ref{22}) remains
a supersymmetric generalization of the $P(X)$ bosonic theory. It will be helpful in the next section if we analyze this result in more detail. First, note that adding (\ref{22}) to the {\it second} term in (\ref{14}) gives
\bea
&&\bigg[\,\frac{8}{16^2}D \P D \P \Db \Pd \Db \Pd \left(\{D,\Db\}(\P - \Pd)\{D,\Db\}(\Pd - \P)\right) \nn \\
&&- \frac{4}{16^3}D \P D \P \Db \Pd \Db \Pd \left(\{D,\Db\}(\P + \Pd)\{D,\Db\}(\P - \Pd)\right)^2 + \frac{1}{16} D{\Phi}
 D{\Phi} {\bar{D}{\Phi}^{\dagger} {\bar{D}} {\Phi}^{\dagger}} \bigg]
\bigg\vert_{\theta \theta {\bar{\theta}} {\bar{\theta}}, \rm quad} \nn \\
&&=\left[-2(\pt\p)^4(\pt\chi)^2 - (\pt\p)^4(\pt\p\cdot\pt\chi)^2\right]+\left[- \frac{1}{2}(\partial\phi)^2(\partial\chi)^2 + (\partial\phi \cdot \partial\chi)^2\right]\,,
 \label{26}
\eea
where we have not shown irrelevant pure $\phi$ terms or terms involving fermions. When evaluated around the ghost condensate vacuum (\ref{16}) with $c=1$,~(\ref{26}) reduces to
\begin{equation}
\left[-2(\pt\chi)^2 - (\dot{\chi})^2\right]+\left[ \frac{1}{2}(\partial\chi)^2 + (\dot{\chi})^2\right]=-\frac{3}{2}(\partial\chi)^2 \ .
\label{27}
\end{equation}
That is, adding (\ref{22}) to the second term of (\ref{14}) exactly cancels the Lorentz-violating term. In addition, the signs are such that the resulting Lorentz-covariant kinetic term for $\chi$ is ghost free with correct sign spatial gradient. Second, adding this to the {\it first} term in (\ref{14}) produces the canonical normalization while leaving the correct sign unchanged. That is,
\begin{equation}
\left[\frac{1}{2}(\pt\chi)^2\right]+\left[-\frac{3}{2}(\partial\chi)^2\right]=-(\partial\chi)^2 \ ,
\label{28}
\end{equation}
which gives (\ref{24}) precisely.

Finally, consider the kinetic term for the fermion fluctuation $\delta \psi$. This is given in the third line of (\ref{20}) and, as with $\delta \chi$, is new to the supersymmetric theory.
We see from (\ref{20}) that, although the magnitudes of the coefficients of the two $\delta \psi$ terms are equal, the time-derivative term is ghost-free while the spatial gradient term has the wrong sign.
Note that this is not the same kind of gradient instability as occurs for $\p$. There, the coefficient of the spatial derivative term is zero or small and, hence, higher-derivative terms can play a role in guaranteeing stability over an extended time period. For $\psi$, on the other hand, the coefficient of the wrong-sign spatial gradient term is not small. It follows that the inclusion of higher-derivative terms, such as those in (\ref{21}), is necessarily irrelevant. The situation for the fermion, therefore, is more akin to that of the second scalar $\chi$, whose deep wrong-sign spatial gradient had to be corrected by the addition of a new second order term --- the sum of the two kinetic spatial gradients having the correct sign.
However, within the context of the supersymmetric extension of the {\it pure $P(X)$ theory}, we are unable to find a fermionic analog of this mechanism. That is, the fermion kinetic {\it spatial gradient} term has the wrong sign!

As discussed in~\cite{Khoury:2010gb}, it is unclear whether or not this is physically unacceptable. This will be explored elsewhere \cite{toappear}. In this paper, we ask a different question: by modifying the bosonic theory so that it is no longer purely a $P(X)$ theory, can one find a supersymmetric extension that is free of both ghost-like and gradient-like instabilities in {\it all} of its component fields? The answer, as we will see, is yes, and leads to another interesting class of higher-derivative Lagrangians --- the conformal Galileon theories.

\section{Curing the Fermion Gradient Instability}
\label{fermioncure}

To solve the gradient instability problem for the fermion, we proceed by analogy with the $\chi$ scalar. That is, 1) we find a supersymmetric interaction which, when added to the second term in (\ref{14}), cancels the Lorentz-violating part of its fermion quadratic terms --- rendering the fermion kinetic term Lorentz-covariant with the correct sign --- and 2) we add this to the first term in (\ref{14}) to canonically normalize the coefficient. However, there is one {\it important caveat}. As stated above, our attempts to do this with precisely the two terms in (\ref{14}) failed. To solve this problem, it  turns out that one must make a {\it mild modification} of each of these terms --- a modification that, however, does not reduce to the pure $P(X)$ theory, or even the generalized $P(X,\phi)$ theory discussed in \cite{Khoury:2010gb}.

With this in mind, recall from (\ref{14}) and (\ref{15}) that
\begin{eqnarray}
\frac{1}{16} D{\Phi}
 D{\Phi} {\bar{D}{\Phi}^{\dagger} {\bar{D}} {\Phi}^{\dagger}}
\Big\vert_{\theta \theta {\bar{\theta}} {\bar{\theta}}}&=&\frac{1}{4}(\partial\phi)^4
-\frac{i}{4}(\partial\phi)^2(\psi_{,\mu}\sigma^\mu {\bar{\psi}} - \psi \sigma^\mu {\bar{\psi}}_{,\mu}) \nn \\
&-&\frac{i}{2} \phi_{,\mu}\phi_{,\nu}(\psi^{,\nu}\sigma^\mu {\bar{\psi}} - \psi \sigma^\mu {\bar{\psi}}^{,\nu})+\dots
\label{29}
\end{eqnarray}
Here and henceforth in this section, we drop irrelevant terms containing $\chi$ and set $F=0$. Let us now modify this term to
\begin{eqnarray}
\left[\frac{1}{4(\Phi+{\Phi}^{\dagger})^{4}} D{\Phi}
 D{\Phi} {\bar{D}{\Phi}^{\dagger} {\bar{D}} {\Phi}^{\dagger}}\right]
\Bigg\vert_{\theta \theta {\bar{\theta}} {\bar{\theta}}}&=&\frac{1}{4\phi^{4}}(\partial\phi)^4-\frac{i}{4\phi^{4}}(\partial\phi)^2(\psi_{,\mu}\sigma^\mu {\bar{\psi}} - \psi \sigma^\mu {\bar{\psi}}_{,\mu}) \nn \\
&-& \frac{i}{2\phi^{4}}\phi_{,\mu}\phi_{,\nu}(\psi^{,\nu}\sigma^\mu {\bar{\psi}} - \psi \sigma^\mu {\bar{\psi}}^{,\nu})+\dots
\label{30}
\end{eqnarray}
To the order we are working, the only effect of $(\Phi+{\Phi}^{\dagger})^{-4}$ is to multiply expression (\ref{29}) by an overall factor of $\phi^{-4}$. Furthermore, setting $\psi=0$  reduces
(\ref{30}) to $X^2/\phi^4$. In other words, this modified term {\it is} a supersymmetric extension of the $P(X,\phi)$ theories discussed in~\cite{Khoury:2010gb}. When evaluated on a ghost condensate background, the first fermionic term in (\ref{30}) remains Lorentz-covariant, while the last term explicitly breaks Lorentz invariance.

Can one find a supersymetric interaction that will exactly cancel this Lorentz-violating fermion kinetic term? Consider
 \bea
 \bigg[\frac{-1}{24(\P+\Pd)^3}\left(D\P D\P \Db^2\Pd + {\rm h.c.}\right)\bigg]\bigg\vert_{\th\th\tb\tb} &=& -\frac{1}{6\p^{3}}\Box\p(\pt\p)^2 \nn \\
 - \frac{i}{6\p^3}\p_{,\m}(\psi_{,\n}\s^\n \bar\psi^{,\m}-\psi^{,\m}\s^\n\pb_{,\n})& + &\frac{i}{12\p^3}\Box \p (\psi_{,\n}\s^\n\pb-\psi\s^\n\pb_{,\n}) \nn \\
- \frac{i}{12\p^3}\p_{,\m}(\psi\s^\m\Box \pb-\Box \psi\s^\m\pb) &-& \frac{i}{4\p^4}(\pt\p)^2(\psi_{,\n}\s^\n\pb-\psi\s^\n\pb_{,\n})\,,
\label{31}
 \eea
where we work to quadratic order in the $\psi$ fluctuations. (Useful intermediate steps in evaluating the above expression can be found in Appendix A). An important technical fact is that, while the first four terms are contained in the component expansion of
$(D\P D\P \Db^2\Pd + {\rm h.c.})\mid_{\th\th\tb\tb}$,
the last term arises due to a contribution from the prefactor. This did not occur in (\ref{30}) which, to the order that concerns us, was simply multiplied by a factor of $\phi^{-4}$. Here, however, the prefactor is {\it significant} and must be included to solve the fermion gradient instability problem.
Integrating the second and fourth terms by parts, and dropping all interactions that vanish on a ghost condensate background, we find that (\ref{31}) dramatically simplifies to
\bea
\bigg[\frac{-1}{24(\P+\Pd)^3}\left(D\P D\P \Db^2\Pd + {\rm h.c.}\right)\bigg]\bigg\vert_{\th\th\tb\tb} &=& -\frac{1}{6\p^{3}}\Box\p(\pt\p)^2\nn \\
&+&\frac{i}{2\p^4}\p_{,\m}\p_{,\n}(\psi^{,\n}\s^\m\pb-\psi\s^\m\pb^{,\n})\,.
\label{32}
\eea
Note that the fermion term is simply $-1$ times the Lorentz-violating last term of (\ref{30}) --- a fact requiring, amongst other things, the $\P+\Pd$  prefactors in both (\ref{30}) and (\ref{32}). Also,
when setting the fermion to zero (\ref{32}) reduces to $ -\Box\p(\pt\p)^2/6\p^{3}$, which is manifestly {\it not} of the $P(X,\phi)$ form. Instead, we recognize this as the cubic term of Galileon
theories (more precisely, conformal Galileon theories, as we will see shortly). The fact that~(\ref{32}) goes beyond the $P(X,\phi)$ form is consistent with our earlier conclusion that the fermionic instability
could not be removed within the context of supersymmetric ghost condensates!

Adding (\ref{30}) and (\ref{32}) together, the Lorentz-violating fermion term {\it exactly cancels} and one obtains the Lorentz-covariant fermionic Lagrangian
\bea
&&\bigg [\frac{-1}{24(\P+\Pd)^3}\left(D\P D\P \Db^2\Pd + {\rm h.c.}\right)+ \frac{1}{4(\Phi+{\Phi}^{\dagger})^{4}} D{\Phi}
 D{\Phi} {\bar{D}{\Phi}^{\dagger} {\bar{D}} {\Phi}^{\dagger}} \bigg] \bigg\vert_{\theta \theta {\bar{\theta}} {\bar{\theta}}} \nn \\
&&= -\frac{1}{6\p^{3}}\Box\p(\pt\p)^2 +\left[\frac{1}{4\phi^{4}}(\partial\phi)^4-\frac{i}{4\phi^{4}}(\partial\phi)^2(\psi_{,\mu}\sigma^\mu {\bar{\psi}} - \psi \sigma^\mu {\bar{\psi}}_{,\mu})\right] \label{33} +\ldots
\eea
Integrating twice by parts, the first term can be expressed as
\begin{equation}
-\frac{1}{6\p^{3}}\Box\p(\pt\p)^2=-\frac{1}{6\phi^{4}}(\partial \phi)^{4}+\frac{1}{18 \phi^{2}}(\partial_{\mu}\partial_{\nu} \phi)^{2}-\frac{1}{18 \phi^{2}}(\Box\phi)^{2} \ .
\label{34}
\end{equation}
It follows that
\bea
&&\bigg [\frac{-1}{24(\P+\Pd)^3}\left(D\P D\P \Db^2\Pd + {\rm h.c.}\right)+ \frac{1}{4(\Phi+{\Phi}^{\dagger})^{4}} D{\Phi}
 D{\Phi} {\bar{D}{\Phi}^{\dagger} {\bar{D}} {\Phi}^{\dagger}} \bigg] \bigg\vert_{\theta \theta {\bar{\theta}} {\bar{\theta}}} \nn \\
 && = \frac{1}{12\phi^{4}}(\partial\phi)^4 +\frac{1}{18 \phi^{2}}(\partial_{\mu}\partial_{\nu} \phi)^{2}-\frac{1}{18 \phi^{2}}(\Box\phi)^{2}  -\frac{i}{4\phi^{4}}(\partial\phi)^2(\psi_{,\mu}\sigma^\mu {\bar{\psi}} - \psi \sigma^\mu {\bar{\psi}}_{,\mu})+\ldots
 \label{35}
\eea
Three fundamental conclusions can be drawn from~(\ref{35}): 1) the fermion kinetic term is Lorentz-covariant and, for any purely time-dependent background, of the correct sign --- that is, {\it ghost-free with correct-sign spatial gradient}; 2) the first term is simply $X^{2}/3\phi^{4}$ and is manifestly of the $P(X,\phi)$ type; 3) the remaining $\phi$ terms
are of a different differential form and {\it not} of the $P(X,\phi)$ type. Thus, by moving away from purely $P(X,\phi)$ theory we have solved the problem of the fermion gradient instability.

As with $\chi$, one must now add this equation to the first term in Lagrangian (\ref{14}). Since canceling the Lorentz-violating fermion kinetic term required a modification of the higher-derivative operators, we must also appropriately modify the first term in (\ref{14}). Clearly, this requires multiplying the $\theta\theta{\bar{\theta}}{\bar{\theta}}$-component of $-\Phi\Phi^{\dagger}$ by $1/\phi^{4}$. Although naively one might think this would be accomplished by the expression $-4\Phi \Phi^{\dagger}/(\P+\Pd)^4$, the correct result is more subtle, as discussed in~\cite{Wess:1992cp}.
Defining
\begin {equation}
K(\Phi,{\Phi}^{\dagger})=\frac{2}{3(\Phi+\Phi^{\dagger})^{2}} \ ,
\label{36}
\end{equation}
the appropriate modification is given by
\begin{equation}
-K(\Phi,{\Phi}^{\dagger})\Big\vert_{\theta \theta {\bar{\theta}} {\bar{\theta}}}=\frac{1}{2\phi^{4}}(\pt\phi)^{2}-\frac{i}{2\phi^{4}}(\psi_{,\mu}\sigma^\mu {\bar{\psi}} - \psi \sigma^\mu {\bar{\psi}}_{,\mu})  \,,
\label{37a}
\end{equation}
where we suppress irrelevant $\chi$ and $F$ contributions.
The first term is just $-X/\phi^{4}$ and hence of the $P(X\phi)$ form.
Although not strictly necessary, we choose to add (\ref{35}) to (\ref{37a}) in such a way that the $X$ dependent contribution for $\phi$ takes the canonical ghost condensate form (\ref{10}). This will be the case if one takes  the complete Lagrangian to be (\ref{37a})+$3\times$(\ref{35}):
\bea
&&\bigg[-K(\Phi,{\Phi}^{\dagger}) - \frac{1}{8(\P+\Pd)^3}\left(D\P D\P \Db^2\Pd + {\rm h.c.}\right) +\frac{3}{4(\Phi+{\Phi}^{\dagger})^{4}} D{\Phi}
 D{\Phi} {\bar{D}{\Phi}^{\dagger} {\bar{D}} {\Phi}^{\dagger}} \bigg] \bigg\vert_{\theta \theta {\bar{\theta}} {\bar{\theta}}} \nn \\
&& = \frac{1}{2\phi^{4}}(\pt\phi)^{2}+\frac{1}{4\phi^{4}}(\pt\phi)^{4}
+\frac{1}{6 \phi^{2}}(\partial_{\mu}\partial_{\nu} \phi)^{2}-\frac{1}{6 \phi^{2}}(\Box\phi)^{2}  \nn \\
&&-\frac{i}{2\phi^{4}}\left(1+\frac{3}{2}(\partial \phi)^{2}\right)(\psi_{,\mu}\sigma^\mu {\bar{\psi}} - \psi \sigma^\mu {\bar{\psi}}_{,\mu}) +\ldots\nn \\
&&=\frac{1}{\phi^{4}}(-X+X^{2})
+\frac{1}{6 \phi^{2}}\left((\partial_{\mu}\partial_{\nu} \phi)^{2}-(\Box\phi)^{2}\right)
+\frac{i}{4\phi^{4}}\left(\psi_{,\mu}\sigma^\mu {\bar{\psi}} - \psi \sigma^\mu {\bar{\psi}}_{,\mu}\right) +\ldots
 \label{38}
\eea

The first bracketed term is of the $P(X,\phi)$ type, whereas the second group of scalar terms is not. Be that as it may, the ghost condensate $\phi=c\,t$ with $c=1$ is a vacuum solution of the equations of motion. The coefficient of the fermion kinetic term in the final line of (\ref{38}) has been evaluated in this vacuum. As promised, the fermion kinetic term is ghost free with correct-sign spatial gradients
 and, rescaling $\psi \rightarrow \sqrt{2}\psi$, has {\it canonical} normalization. By canonical we mean that the ratio of the fermion kinetic coefficient to the coefficients of the $\phi$ and $\chi$ kinetic terms is the same as in a standard supersymmetric Lagrangian. Note, however, that in the present case all three kinetic terms are multiplied by the common prefactor $1/\phi^{4}=1/t^{4}$.
For completeness, we point out that similar vacua can be achieved by choosing the Lagrangian to be
$-c_{2}\times$(\ref{37a})$-c_{3}\times 3 \times $(\ref{35}) for any negative coefficients $c_{2},c_{3}${\footnote{The coefficients $c_{i}$ are defined so as to conform with the standard notation used in the following section.}}. In this case, a ghost condensate solution $\phi=c\,t$ still exists, but with $c=\sqrt{c_{2}/c_{3}}$.

Although the equations of motion continue to have a ghost condensate solution of the form (\ref{16}), the Lagrangian (\ref{38}) is {\it not} a supersymmetrized ghost condensate theory! Therefore, the price one pays to solve the gradient instability problem for the fermion is a modification of the bosonic part of the theory. Remarkably, (\ref{38}), and its generalizations to arbitrary positive coefficients $-c_{2}$ and $-c_{3}$, is precisely the Lagrangian for a well-known class of higher-derivative models --- the {\it conformal Galileon theories} --- to which we now turn.

\section{Galileons and their Relation to $P(X,\phi)$ Theories}
\label{galP(X)}

Galileon scalar field theories were first discovered in the context of the Dvali-Gabadadze-Porrati (DGP) brane-world model~\cite{Dvali:2000hr,Dvali:2000xg} and arise generically in brane-induced gravity models~\cite{Dvali:2002pe}$-$\cite{deRham:2010rw}. In a certain decoupling limit~\cite{Deffayet:2001uk,Luty:2003vm,ArkaniHamed:2002sp}, the theory becomes local in four dimensions
and describes a real scalar field $\pi$ (the brane-bending mode) with
\be
{\cal L}_{\rm DGP} = -\frac{1}{2}(\partial\pi)^2 - \frac{1}{\Lambda^3}(\partial\pi)^2\Box\pi + \frac{1}{\sqrt{6}M_{\rm Pl}} \pi T^\mu_{\;\mu}\, ,
\label{LDGP}
\ee
where $T_{\mu\nu}$ is an external source, and the strong coupling scale $\Lambda$ is related to the four- and five-dimensional Planck scales as $\Lambda \equiv \sqrt{6}M_5^2/M_{\rm Pl}$. (Despite the conformal coupling to $T^\mu_{\;\mu}$, the theory is nevertheless consistent with tests of gravity because $\pi$ is screened in the vicinity of massive sources~\cite{Deffayet:2001uk,Dvali:2007kt,ArkaniHamed:2002sp,Vainshtein:1972sx}. See~\cite{Jain:2010ka,Khoury:2010xi} for reviews of screening mechanisms.) As a vestige of five-dimensional Poincar\'e invariance, this theory has two independent internal symmetries~\cite{Luty:2003vm,Nicolis:2004qq},
\bea
&& \delta_{c}\pi=c \nn \\
&& \delta_{v}\pi=v_{\mu}x^{\mu}\,,
\label{galshift}
\eea
where $c$ and $v^\mu$ are constant. The first transformation in~(\ref{galshift}) is just a standard shift symmetry, whereas the second is called a Galilean symmetry.
The latter protects the cubic interaction from being renormalized~\cite{Luty:2003vm,Nicolis:2004qq}. Remarkably, despite its higher-derivative form,~(\ref{LDGP}) leads to an equation of motion that is {\it second-order} in derivatives.

In~\cite{Nicolis:2008in}, Lagrangian~(\ref{LDGP}) was generalized to include all possible interactions that are invariant under the shift and Galilean symmetries, and which lead to second-order equations of motion. In addition to the linear, quadratic and cubic terms in $\pi$ shown in~(\ref{LDGP}), it was found that quartic and quintic interactions are also allowed. The most general ``Galileon'' theory is found to be a linear combination of the Lagrangians~\cite{Nicolis:2008in}
\bea
{\cal L}_{{\rm Gal},\,2} &=& -\frac{1}{2} (\pt \pi)^2 \nn\\
 {\cal L}_{{\rm Gal},\,3} &=& -\frac{1}{2}(\pt\pi)^2\Box\pi  \nn \\
{\cal L}_{{\rm Gal},\,4} &=& (\pt\pi)^2\left[-\frac{1}{2}(\Box\pi)^2+\frac{1}{2}\pi^{,\mu\nu}\pi_{,\mu\nu}\right] \nn \\
 {\cal L}_{{\rm Gal},\,5} &=& (\pt\pi)^2\left[-\frac{1}{2}(\Box\pi)^3 -\pi^{,\mu\nu}\pi_{,\nu\rho}{\pi^{,\rho}}_\mu + \frac{3}{2}\Box\pi\pi^{,\mu\nu}\pi_{,\mu\nu}\right]\, ,
 \label{dude}
 \eea
where we have set the associated mass scales of each term to unity to simplify notation.
As with the cubic term, these interactions are protected by non-renormalization theorems. The construction stops with $ {\cal L}_{{\rm Gal},\,5}$ --- no higher-order interactions can satisfy the simultaneous requirements of shift/Galilean invariance and second-order equations.

The symmetries in~(\ref{galshift}) can be promoted to a subgroup of the conformal group, with infinitesimal dilation and special conformal transformations acting respectively as
 \bea
 && \delta_{c}\pi=c\left(1+x^{\mu}\pt_{\mu}\pi\right) \nn \\
 && \delta_{v}\pi=v_{\mu}x^{\mu}-\pt_{\mu}\pi\left(\frac{1}{2}v^{\mu}x^{2}-(v \cdot x)x^{\mu}\right) \,.
  \label{42a}
 \eea
 In the limit of small $\pi$, these reduce to~(\ref{galshift}). The unique Lagrangians invariant under these symmetries and leading to second-order equations of motion are~\cite{Nicolis:2008in,deRham:2010eu}
 \bea
 {\cal L}_2 &=& -\frac{1}{2} e^{2\pi}(\pt \pi)^2 \nn \\
 {\cal L}_3 &=& -\frac{1}{2}(\pt\pi)^2\Box\pi - \frac{1}{4}(\pt\pi)^4 \nn \\
 {\cal L}_4 &=& e^{-2\pi}(\pt\pi)^2\bigg[-\frac{1}{2}(\Box\pi)^2+\frac{1}{2}\pi^{,\mu\nu}\pi_{,\mu\nu} \nn \\
&&+ \frac{1}{5}(\pt\pi)^2\Box\pi-\frac{1}{5}\pi^{,\mu}\pi^{,\nu}\pi_{,\mu\nu} -\frac{3}{20}(\pt\pi)^4\bigg] \nn \\
 {\cal L}_5 &=& e^{-4\pi}(\pt\pi)^2\bigg[-\frac{1}{2}(\Box\pi)^3 -\pi^{,\mu\nu}\pi_{,\nu\rho}{\pi^{,\rho}}_\mu \nn \\
 &&+\frac{3}{2}\Box\pi\pi^{,\mu\nu}\pi_{,\mu\nu}+\frac{3}{2}(\pt\pi)^2(\Box\pi)^2
-\frac{3}{2}(\pt\pi)^2\pi^{,\mu\nu}\pi_{,\mu\nu}  \nn \\
&&-\frac{15}{7}(\pt\pi)^4\Box\pi+\frac{15}{7}(\pt\pi)^2\pi^{,\mu}\pi^{,\nu}\pi_{,\mu\nu} -\frac{3}{56}(\pt\pi)^6\bigg ]\,.
\label{42}
 \eea
 The class of theories obtained by taking general linear combinations of these terms are called {\it conformal} Galileon theories.
To compare Galileons to $P(X,\p)$ theories, it is useful to change variables to
 \be \phi \equiv e^{-\pi} \ .
 \label{43}
 \ee
The above Lagrangians then become
\bea
{\cal L}_2 &=& -\frac{1}{2\p^4}(\pt \p)^2 \nn \\
{\cal L}_3 &=& \frac{1}{2\p^3}\Box\p(\pt\p)^2 -\frac{3}{4\p^4}(\pt\p)^4\nn \\ &=& -\frac{1}{4\phi^{4}}(\pt\phi)^{4}-\frac{1}{6 \phi^{2}}(\partial_{\mu}\partial_{\nu} \phi)^{2}+\frac{1}{6 \phi^{2}}(\Box\phi)^{2}\nn \\
{\cal L}_4 &=& -\frac{1}{2\p^2}(\pt\p)^2(\Box\p)^2 +\frac{1}{2\p^2}(\pt\p)^2\p^{,\mu\nu}\p_{,\mu\nu} +\frac{4}{5\p^3}(\pt\p)^4\Box\p\nn \\
&& -\frac{4}{5\p^3}(\pt\p)^2\p^{,\m}\p^{,\nu}\p_{,\mu\nu}-\frac{3}{20\p^4}(\pt\p)^6\nn  \\
{\cal{L}}_{5} &=&  (\pt\p)^2\bigg[\frac{1}{2\p}(\Box\p)^3 + \frac{1}{\p} \p^{,\mu\nu}\p_{,\nu\rho}{\p^{,\rho}}_\mu \nn \\
 &&-\frac{3}{2\p}\Box\p\p^{,\mu\nu}\p_{,\mu\nu}-\frac{3}{4\p^2}\pt_\mu(\pt\p)^2\pt^\mu(\pt\p)^2
+\frac{3}{\p^2}\Box\p \p^{,\mu\nu}\p_{,\mu}\p_{,\nu}  \nn \\
&&+\frac{6}{7\p^3}(\pt\p)^2 \p^{,\mu\nu}\p_{,\mu}\p_{,\nu} -\frac{6}{7\p^3}(\pt\p)^4\Box\p -\frac{3}{56\p^4}(\pt\p)^8\bigg ]\nn \\  &=&  (\pt\p)^2\bigg[\frac{1}{2\p}(\Box\p)^3 + \frac{1}{\p} \p^{,\mu\nu}\p_{,\nu\rho}{\p^{,\rho}}_\mu \nn \\
 &&-\frac{3}{2\p}\Box\p\p^{,\mu\nu}\p_{,\mu\nu}-\frac{3}{4\p^2}(\pt\p)^2(\Box\p)^2
+\frac{3}{4\p^2}(\pt\p)^2\p^{,\mu\nu}\p_{,\mu\nu}  \nn \\
&&+\frac{9}{14\p^3}(\pt\p)^4\Box\p-\frac{9}{14\p^3}(\pt\p)^2\p^{,\mu}\p^{,\nu}\p_{,\mu\nu} -\frac{3}{56\p^4}(\pt\p)^8\bigg ] \,,
 \label{45}
\eea
where the second versions of ${\cal L}_3$ and ${\cal L}_5$ follow from integration by parts.

Note that although the bosonic Galileon Lagrangians ${\cal {L}}_2$ and ${\cal {L}}_3$ were introduced for entirely different reasons, they are {\it precisely} of the form --- derived in detail in Sec.~3 --- required by a quadratic and cubic supersymmetric theory to have a ghost condensate vacuum with {\it Lorentz-covariant} and {\it canonical sign} fermion kinetic energy. Specifically, the purely $\phi$-dependent part of (\ref{37a}) and $3 \times (\ref{35})$ are
\begin{equation}
 \frac{1}{2\p^4}(\pt \p)^2=-{\cal L}_2
 \label{47a}
 \end{equation}
 and
 \begin{equation}
 \frac{1}{4\phi^{4}}(\pt\phi)^{4}+\frac{1}{6 \phi^{2}}(\partial_{\mu}\partial_{\nu} \phi)^{2}-\frac{1}{6 \phi^{2}}(\Box\phi)^{2} = -{\cal L}_3  \,,
 \label{47b}
 \end{equation}
respectively! Thus, as claimed in Sec.~3, Galileon theories arise naturally in generalized supersymmetric ghost condensate theories, independently of their original origin in \cite{Dvali:2000hr,Luty:2003vm,Nicolis:2004qq,Nicolis:2008in}.
It is of interest, although somewhat peripheral to our main discussion, to ask what scalar sector would emerge if we allowed the quadratic and cubic supersymmetrized ghost condensate theory to have a Lorentz-violating fermion kinetic term. This possibility is explored in detail in Appendix B.

The most general Galileon Lagrangian is given by
\be
{\cal L} = c_2 {\cal L}_2 + c_3 {\cal L}_3 +c_4 {\cal L}_4 + c_5 {\cal L}_5\,,
\label{48}
\ee
where the $c_i$ coefficients are constant. Restricting to time-dependent fields only, it follows from (\ref{45}) that
\be
{\cal L} = \frac{1}{\p^4}P(X)\,; \quad P(X)= c_2 X -c_3 X^2 +\frac{6}{5} c_4 X^3 - \frac{6}{7}c_5 X^4\,,
\label{ConformalPofX}
\ee
where $X= \dot\p^2/2$. In other words, {\it for purely time-dependent backgrounds, the Galileon theory reduces to a $P(X)$ theory with an overall multiplicative factor of $\phi^{-4}$}. However, the spatial gradients are radically different, and much better behaved, than in the $P(X,\phi)$ case. This fact has important consequences, which we discuss below. Nevertheless, the connection with $P(X,\phi)$ theories considerably simplifies the analysis of time-dependent solutions, as we now demonstrate.

It follows from the above discussion that the ``de Sitter'' solution
\be \pi_{\rm dS} = -\ln(H_0 t)
\label{50}
\ee
of the Galileon theory is simply the ghost-condensate solution
\be
\p = H_0 t \,
\label{51}
\ee
of the associated $P(X,\p)$ theory. The constant $H_0$, as it is usually denoted in the Galileon literature, thus corresponds to the coefficient $c$ of the ghost condensate,
\be
c = H_0\,.
\ee
For such solutions, where $X$ is constant, the equation of motion derived from action (\ref{ConformalPofX}) is
\be
2XP_{,X} -P =0\,.
 \label{equationofmotion}
 \ee
In terms of the $c_i$ coefficients and $H_0$, this reduces to
\be
c_2 -\frac{3}{2}c_3 H_0^2+\frac{3}{2}c_4 H_0^4-\frac{3}{4}c_5 H_0^6=0\,.
\label{53}
 \ee
Interestingly, since the energy density $\rho$ is given by
\bea
\rho &=& \frac{1}{\p^4}(2XP_{,X}-P)\,,
\label{54}
\eea
the equation of motion~(\ref{equationofmotion}) implies that $\rho=0$. That is, in order to have a solution with constant $X$, the energy density must vanish. This fact, which appeared to be coincidental in previous treatments of the Galileon self-accelerating solution, can now be seen to be a general requirement.

What are the conditions for these $X={\rm const.}$ solutions to be stable against small perturbations? As demonstrated in~\cite{Nicolis:2008in}, Galileon theories have the property that they modify the spatial gradients of the $\p$ field in just such a way as to render the Lagrangian for fluctuations $\d\p(t,\vec{x})$ covariant, despite the Lorentz-breaking background. Hence, it suffices to require positivity of the temporal kinetic term of the perturbations. Substituting
\be
\p = H_0 t + \d\p(t,\vec{x})
\label{55}
\ee
into (\ref{ConformalPofX}) yields the quadratic Lagrangian
\be
{\cal L}_{\rm{quadratic}} = \frac{1}{H_0^4 t^4} (\dot{\d\p})^2 \left(\frac{1}{2} c_2 - \frac{3}{2}c_3 H_0^2 + \frac{9}{4} c_4 H_0^4 - \frac{3}{2} c_5 H_0^6\right) + \ldots
\label{56}
\ee
Stability requires the expression in brackets to be positive; that is,
\be
c_2 - 3 c_3 H_0^2 + \frac{9}{2} c_4 H_0^4 - 3 c_5 H_0^6 > 0 \ .
\label{57}
\ee
Note that the this inequality can easily be satisfied simultaneously with constraint~(\ref{53}) derived from the equation of motion.

\section{Violating the Null Energy Condition}
\label{NECviolate}

Galileons are interesting theoretically because they can violate the NEC
\be
T_{\mu\nu} n^\mu n^\nu \geq 0\,,
\label{NECcov}
\ee
where $n^\mu$ is an arbitrary null vector, while having {\it stable} temporal and spatial fluctuations. This is no small feat since, under very general conditions, theories with two derivatives are inevitably plagued with ghost or gradient instabilities on NEC-violating backgrounds~\cite{Dubovsky:2005xd}. Galileon (and ghost condensate~\cite{Creminelli:2006xe}) Lagrangians circumvent this problem by having more than two derivatives and, hence, can have vacua with stable violations of the NEC~\cite{Nicolis:2009qm}. This is particularly interesting for cosmological applications. In a cosmological context,~(\ref{NECcov}) reduces to $\rho + {\cal{P}} \geq 0,$ where ${\cal P}$ denotes pressure. Since $\dot{H} = -(\rho + {\cal{P}})/2$, violating this inequality then allows the universe to bounce from a contracting to an expanding phase.

In this section, we derive the conditions under which the conformal Galileon Lagrangian (\ref{48}) violates the NEC.
Although this question has been studied in earlier work~\cite{Nicolis:2009qm}, here we point out important {\it new} ambiguities in defining the
stress tensor, above and beyond the usual field theory ambiguities in $T_{\mu\nu}$. Remarkably, even on a flat-space
background (which is what we consider in this paper), $T_{\mu\nu}$ is sensitive to how one defines the theory, {\it including total
divergence terms}. In particular, two flat-space Lagrangians that differ only by integration by parts can have physically different stress tensors.

Before proceeding, we point out an important distinction between
Galileons and ghost condensate theories. Recall from Section~\ref{galP(X)} that for purely time-dependent solutions, $\phi = \phi(t)$, the conformal Galileon Lagrangian~(\ref{48})
reduces to the particular $P(X,\phi)$ theory in ~(\ref{ConformalPofX}). The full Galileon Lagrangian, of course, differs from the corresponding $P(X,\phi)$ theory
by spatial gradient terms, but these are irrelevant in computing the $\phi$ equation of motion~(\ref{53}), its energy density~(\ref{54})
and the ghost-free condition~(\ref{57}). However, the gradient terms {\it are} important for computing the pressure. Setting the gradient terms to zero in the action does not commute with varying the action to obtain the pressure, as we will see explicitly in the examples below.

A standard way to derive the stress tensor is by covariantizing the theory and varying with respect to the metric. Alternatively,
entirely within the context of field theory on flat space, the stress tensor is derived via the Noether/Belinfante method. For theories that include
up to two derivatives per field, that is, ${\cal L} (\phi,\partial_\mu\phi,\partial_\mu\partial_\nu\phi)$,
the Belinfante stress tensor is given by~\cite{Guarrera:2007tu}
\bea
\nonumber
&& T^{\mu\nu} = \eta^{\mu\nu}{\cal L} - \partial_\lambda\left(\frac{\partial {\cal L}}{\partial (\partial_\mu\partial_\nu \phi)}\partial^\lambda\phi\right) \\
&&- \frac{1}{2}\frac{\partial {\cal L}}{\partial (\partial_\mu\phi)}\partial^{\nu}\phi - \frac{1}{2}\frac{\partial {\cal L}}{\partial (\partial_\nu\phi)}\partial^{\mu}\phi + \partial_\lambda \left(\frac{\partial {\cal L}}{\partial (\partial_\lambda\partial_{\mu} \phi)}\right)\partial^{\nu}\phi + \partial_\lambda  \left(\frac{\partial {\cal L}}{\partial (\partial_\lambda\partial_{\nu} \phi)}\right)\partial^{\mu}\phi\,.\nn\\
\label{belin}
\eea
By construction, this is both symmetric and conserved. Moreover, in all cases we have checked, the Belinfante definition
agrees with the covariantization method and, hence, gives the correct stress tensor that sources gravity.

To compute the pressure, let us set $\mu = \nu = i$ and assume $\phi=\phi(t)$. In this case, the second line in~(\ref{belin}) vanishes and the pressure is given by
\be
{\cal P} = {\cal L} + \frac{{\rm d}}{{\rm d}t} \left(\frac{\partial {\cal L}}{\partial (\partial_i\partial_i \phi)}\dot{\phi}\right)\,,
\label{pressure}
\ee
where no summation is assumed in the second term. This clearly elucidates the difference between
Galileons and the corresponding $P(X,\phi)$ theories mentioned earlier. For pure $P(X,\phi)$, the second term is
manifestly absent, and~(\ref{pressure}) reduces to the usual ${\cal P} = {\cal L} = P(X,\phi)$. For Galileons, however,
the second term will in general contribute to the pressure, even on purely time-dependent backgrounds.

First consider ${\cal L}_2 = -(\partial\phi)^2/2\phi^4$. In this case,~(\ref{pressure}) gives
\be
{\cal P}_2 = \frac{1}{2\phi^4} \dot{\phi}^2 = \frac{1}{2H_0^2t^4} \,,
\ee
where we have substituted~(\ref{51}) in the last step. Next, let us compute the pressure for ${\cal L}_3$.
This is the simplest example that displays the integration-by-parts ambiguities alluded to earlier.
We begin with the definition of ${\cal L}_3$ given by the second line in~(\ref{45}),
\be
{\cal L}_3^{1{\rm st}\;{\rm version}} = \frac{1}{2\p^3}\Box\p(\pt\p)^2 -\frac{3}{4\p^4}(\pt\p)^4\,.
\label{L3v1}
\ee
In this case, $\Box\p(\pt\p)^2$ contributes to the second term in~(\ref{pressure}). Hence, the pressure differs from the corresponding $P(X,\phi)$ result and
\be
{\cal P}_3^{1{\rm st}\;{\rm version}}  = \frac{3}{2\phi^4}\dot{\phi}^4 -  \frac{3}{4\phi^4}\dot{\phi}^4 = \frac{3}{4t^4}\,.
\label{Pv1}
\ee
This agrees with ${\cal P}_3$ derived in~\cite{Nicolis:2009qm}, since their definition of ${\cal L}_3$ was identical to~(\ref{L3v1}).

Now, instead, define ${\cal L}_3$ by the third line in~(\ref{45}), namely
\be
{\cal L}_3^{2{\rm nd}\;{\rm version}} =  -\frac{1}{4\phi^{4}}(\pt\phi)^{4}-\frac{1}{6 \phi^{2}}(\partial_{\mu}\partial_{\nu} \phi)^{2}+\frac{1}{6 \phi^{2}}(\Box\phi)^{2}\,.
\label{L3v2}
\ee
Although this was obtained from~(\ref{L3v1}) solely by integration by parts, the associated pressure is different! Indeed,
\be
{\cal P}_3^{2{\rm nd}\;{\rm version}} = -\frac{1}{4t^4}\,,
\label{Pv2}
\ee
which disagrees with~(\ref{Pv1}).
The resolution of this paradox is as follows. As mentioned earlier, the Belinfante stress tensor gives the same answer as the covariantization
method evaluated on a flat background. The point is that although~(\ref{L3v1}) and~(\ref{L3v2}) differ only by a total derivative,
{\it their covariant versions do not}. Indeed, in going from~(\ref{L3v1}) to~(\ref{L3v2}) we have canceled the terms
\be
\frac{1}{6\phi^2}\left(\partial_\mu\Box\phi - \Box\partial_\mu\phi\right)\partial^\mu\phi\,.
\ee
Although fully justified in flat space, such terms do not cancel on a {\it curved background}. Instead, they give rise to the non-minimal coupling
\be
\frac{1}{6\phi^2}\left(\nabla_\mu\Box\phi - \Box\nabla_\mu\phi\right)\partial^\mu\phi = - \frac{1}{6\phi^2}R_{\mu\nu}\partial^\mu\phi\partial^\nu\phi\,.
\ee
Even though this non-minimal coupling vanishes on a flat background, its variation does not! It is the contribution of the variation of this non-minimal term to the stress tensor that accounts for the
discrepancy between~(\ref{Pv1}) and~(\ref{Pv2}). The lesson is that the stress tensor of Galileon theories, thanks to their higher-derivative nature,
depends on the precise form of the theory in flat space. If two Lagrangians differ by a total derivative, then their stress tensors will agree provided that
in the process of integrating by parts one only cancels terms that would also cancel on a curved background. For example, the stress tensor for ${\cal L}_2$ is unambiguous, but that of ${\cal L}_3$
and higher-order Galileon terms clearly are not.

In the next section, we will see that in order to supersymmetrize ${\cal L}_4,$ it is most convenient to use a new version of ${\cal L}_4,$ related to the version in (\ref{45}) by integration by parts. In order to avoid any confusion, we quote the results for the pressure for the conformal Galileon Lagrangians, as defined in~(\ref{42}), and then also for the version that we supersymmetrize. With the Galileon Lagrangians defined as in~(\ref{42}), substituting into~(\ref{belin}) and setting $\pi(t) = -\ln(H_0t)$ as the background solution, the pressure is
\bea
{\cal P}_2 &=& \frac{1}{2H_0^2t^4} \nn \\
{\cal P}_3 &=& \frac{3}{4t^4} \nn \\
{\cal P}_4 &=& -\frac{9H_0^2}{4t^4} \label{dude2} \\
{\cal P}_5 &=& -\frac{21H_0^4}{8t^4}\,. \nn
\eea
In particular, since our convention for the form of ${\cal L}_4$ and ${\cal L}_5$ differs from that of~\cite{Nicolis:2009qm}, our results for the pressure do not agree.
We have checked that all of these agree with the covariantization method, evaluated on a flat-space background. Since $\rho = 0$ on this background, the condition for violating the NEC for the full Lagrangian~(\ref{48}) is therefore
\be
{\cal P} \propto c_2 + \frac{3}{2}c_3H_0^2 -\frac{9}{2}c_4H_0^4 - \frac{21}{4}c_5H_0^6 < 0\,.
\ee
When we supersymmetrize these theories, we use the same form for ${\cal L}_2$ and ${\cal L}_3,$ but for ${\cal L}_4$ we use instead the last line in (\ref{67}), and, for the reasons described before, we do not supersymmetrize ${\cal L}_5$ explicitly. We then obtain
\be
{\cal P}_4^{\rm SUSY} = \frac{3H_0^2}{4t^4}\,,
\ee
and in our case the condition for violating the NEC becomes
\be
{\cal P}^{\rm SUSY} \propto c_2 + \frac{3}{2}c_3H_0^2 +\frac{3}{2}c_4H_0^4  < 0\,.
\ee

\section{Supersymmetric Galileons}
\label{susygal}

Having discussed Galileons associated with a single real scalar field $\phi$, we proceed to supersymmetrize these theories by embedding $\phi$ in an ${\cal{N}}=1$ chiral superfield $\Phi=(\phi,\chi,\psi,F)$. The procedure we follow is identical to that used in supersymmetrizing $P(X,\phi)$ theories in~\cite{Khoury:2010gb} and employs formulae discussed there, such as the supersymmetry algebra
\be
\{ D_\a , \Db_{\ad} \} = -2i \s^{\mu}_{\a\ad}\pt_{\mu} \label{susyalgebra}
\ee
and its immediate consequence
\be
\{ D,\Db\}\Psi \{ D,\Db\} \Xi = -8 \pt \Psi \pt \Xi
\label{59}
\ee
for any chiral superfields $\Psi$ and $\Xi$.
In addition, in writing the supersymmetric extensions of the Galileon Lagrangians we are making use of several new building blocks, whose component expressions have been written out explicitly in Appendix A.

We find that the possible supersymmetric extensions of ${\cal L}_2$ and ${\cal L}_3$ are very limited. However, there are several options on how to build supersymmetric extensions of the $\p$-dependent Lagrangian ${\cal L}_4$. When we reach ${\cal L}_5,$ the choices on how to supersymmetrize become so numerous, and the corresponding expressions so large, that it becomes impractical --- and not very illuminating --- to write them out explicitly, though there is no obstacle in principle. Hence, we only consider the Galileon Lagrangians up to ${\cal L}_4$ from this point on.

\subsection{${\cal{L}}_{2}$}

The supersymmetric extension of ${\cal{L}}_{2}$ in (\ref{45}) has already been discussed in Sec.~\ref{fermioncure}.
Defining
\begin {equation}
K(\Phi,{\Phi}^{\dagger})=\frac{2}{3(\Phi+\Phi^{\dagger})^{2}} \ ,
\label{60}
\end{equation}
the complete supersymmetrized ${\cal{L}}_{2}$ action is given by
\bea
&&{\cal{L}}_{2}^{\rm SUSY}=K(\Phi,{\Phi}^{\dagger})\Big\vert_{\theta \theta {\bar{\theta}} {\bar{\theta}}}\nn \\
&&=-\frac{1}{2\phi^{4}}(\pt\phi)^{2}-\frac{1}{2\phi^{4}}(\pt\chi)^{2}+\frac{1}{\phi^{4}}F^{*}F +\frac{i}{2\phi^{4}}(\psi_{,\mu}\sigma^\mu {\bar{\psi}} - \psi \sigma^\mu {\bar{\psi}}_{,\mu}) \ .
 \label{61}
\eea
Note that this matches the corresponding expression in~(\ref{45}) when $\chi=F=\psi=0$.

\subsection{${\cal L}_3$}

In Sec.~\ref{fermioncure}, we ``discovered'' the third-order Galileon Lagrangian by looking for a cure for the wrong-sign fermionic spatial gradient term obtained in supersymmetrizing the ordinary ghost condensate theory. We now examine ${\cal L}_3$ in more detail, including all fields of the chiral supermultiplet $(\p,\chi,F,\psi).$ Working to quadratic order in all fields except for $\phi$, we find
\bea
&& \bigg [\frac{1}{(\P+\Pd)^3}\left(D\P D\P \Db^2\Pd + {\rm h.c.}\right)\bigg]\bigg\vert_{\th\th\tb\tb,\,{\rm quad}} = \frac{4}{\p^{3}}\Box\p(\pt\p)^2 -\frac{4}{\p^3}\Box\p(\pt\chi)^2 + \frac{8}{\p^3}\Box\chi (\pt\p\cdot\pt\chi) \nn\\
& & - \frac{8}{\p^3}\Box\p \Fs F + \frac{12}{\p^4}(\pt\p)^2\Fs F  + \frac{4i}{\p^3}\p_{,\m}(\psi_{,\n}\s^\n \bar\psi^{,\m}-\psi^{,\m}\s^\n\pb_{,\n}) - \frac{2i}{\p^3}\Box \p (\psi_{,\n}\s^\n\pb-\psi\s^\n\pb_{,\n})
\nn \\
&& +   \frac{2i}{\p^3}\p_{,\m}(\psi\s^\m\Box \pb-\Box \psi\s^\m\pb) + \frac{6i}{\p^4}(\pt\p)^2(\psi_{,\n}\s^\n\pb-\psi\s^\n\pb_{,\n})
\label{62}
\eea
and
\begin{eqnarray}
&& \bigg[\frac{1}{(\Phi+{\Phi}^{\dagger})^{4}} D{\Phi}  D{\Phi} {\bar{D}{\Phi}^{\dagger} {\bar{D}} {\Phi}^{\dagger}} \bigg]\bigg\vert_{\theta \theta {\bar{\theta}} {\bar{\theta}}, \,{\rm quad}} =\frac{1}{\phi^{4}}(\partial\phi)^4-\frac{2}{\p^4}(\pt\p)^2(\pt\chi)^2 + \frac{4}{\p^4}(\pt\p\cdot\pt\chi)^2  \nn \\ && -\frac{4}{\p^4}(\pt\p)^2\Fs F-\frac{i}{\phi^{4}}(\partial\phi)^2(\psi_{,\mu}\sigma^\mu {\bar{\psi}} - \psi \sigma^\mu {\bar{\psi}}_{,\mu})- \frac{2i}{\phi^{4}}\phi_{,\mu}\phi_{,\nu}(\psi^{,\nu}\sigma^\mu {\bar{\psi}} - \psi \sigma^\mu {\bar{\psi}}^{,\nu})\,.
\label{63}
\end{eqnarray}
These can be combined to give a supersymmetric extension of the ${\cal L}_3$ conformal Galileon Lagrangian
\bea
{\cal L}_{3, \,{\rm quad}}^{\rm SUSY} &=& \frac{1}{8(\P+\Pd)^3}\left[D\P D\P \Db^2\Pd + {\rm h.c.}\right]\Big\vert_{\th\th\tb\tb}-\frac{3}{4(\P + \Pd)^4}D\P D\P \Db\Pd\Db\Pd \Big\vert_{\th\th\tb\tb} \nn \\
&=& -\frac{1}{4\phi^{4}}(\pt\phi)^{4}-\frac{1}{6 \phi^{2}}(\partial_{\mu}\partial_{\nu} \phi)^{2}+\frac{1}{6 \phi^{2}}(\Box\phi)^{2} \nn \\
&& -\frac{1}{\phi^{3}}\pt^{\mu}\chi \pt^{\nu}\chi\pt_{\mu}\pt_{\nu}\phi +\left(-\frac{1}{\phi^{3}}\Box\phi+\frac{9}{2\phi^{4}}(\pt\phi)^{2}\right)F^{*}F \nn \\
&&+ \frac{5i}{\p^4}(\pt\p)^2(\psi_{,\n}\s^\n\pb-\psi\s^\n\pb_{,\n}) - \frac{2i}{\p^3}\Box \p (\psi_{,\n}\s^\n\pb-\psi\s^\n\pb_{,\n})
\nn \\ &&+ \frac{4i}{\p^3}\p_{,\m}(\psi_{,\n}\s^\n \bar\psi^{,\m}-\psi^{,\m}\s^\n\pb_{,\n}) + \frac{2i}{\p^3}\p_{,\m}(\psi\s^\m\Box \pb-\Box \psi\s^\m\pb)  \nn \\
&& - \frac{2i}{\phi^{4}}\phi_{,\mu}\phi_{,\nu}(\psi^{,\nu}\sigma^\mu {\bar{\psi}} - \psi \sigma^\mu {\bar{\psi}}^{,\nu})\,.
\label{64}
\eea
To obtain the bosonic part of (\ref{64}), we have used (\ref{34}) and the fact that, integrating by parts,
\bea
 \frac{1}{\p^3}\Box \chi (\pt\p\cdot\pt\chi) = \frac{3}{\p^3}(\pt\p\cdot\pt\chi)^2 - \frac{1}{\p^3}\chi^{,\mu}\chi^{,\nu}\p_{,\mu\nu}  -\frac{3}{2\p^4}(\pt\p)^2(\pt\chi)^2+\frac{1}{2\p^3}\Box\p (\pt\chi)^2.
\label{65}
\eea
Note that~(\ref{64}) reduces to ${\cal L}_{3}$ in (\ref{45}) when $\chi=\psi=F=0$, as it should.

Finally, integrating the third and fourth fermion terms by parts and dropping any term that vanishes on a ghost condensate background (where $\p_{,\m\n}=0$ and, hence, $X={\rm const.}$), ${\cal L}_{3,\,{\rm quad}}^{\rm SUSY}$ reduces to
\be
{\cal L}^{\rm SUSY}_{3,\, {\rm quad},\, X={\rm const}} = - \frac{1}{4\p^4}(\pt\p)^4 + \frac{9}{2\p^4}(\pt\p)^2 \Fs F + \frac{3i}{4\p^4}(\pt\p)^2(\psi_{,\m}\s^\m \pb - \psi \s^\m \pb_{,\m})\,.
\label{66}
\ee
Remarkably, at quadratic order, the second scalar $\chi$ does not contribute to ${\cal L}_3^{\rm SUSY}$ on a constant $X$ background. Another interesting feature of this Lagrangian is that, despite its higher-derivative nature, no kinetic term for $F$ is generated. Hence, the auxiliary field can be eliminated as described in our previous paper \cite{Khoury:2010gb}.

\subsection{${\cal L}_4$}

While the supersymmetric extension of ${\cal L}_3$ derived above was relatively straightforward, the analogous construction for ${\cal L}_4$ is more complicated. First of all, for ${\cal L}_4$ in~(\ref{45}) the building blocks necessary to constructing supersymmetric generalizations are not manifest.
It is useful, therefore, to use integration by parts to rewrite this fourth-order Lagrangian as
\bea
{\cal L}_4 &=& -\frac{1}{2\p^2}(\pt\p)^2(\Box\p)^2 +\frac{1}{2\p^2}(\pt\p)^2\p^{,\mu\nu}\p_{,\mu\nu} +\frac{4}{5\p^3}(\pt\p)^4\Box\p \nn \\ && -\frac{4}{5\p^3}(\pt\p)^2\p^{,\m}\p^{,\nu}\p_{,\mu\nu} -\frac{3}{20\p^4}(\pt\p)^6 \nn \\ &=& -\frac{1}{4\p^2}\pt_\mu(\pt\p)^2\pt^\mu(\pt\p)^2 +\frac{1}{\p^2}\Box\p\p^{,\mu}\p^{,\nu}\p_{,\mu\nu} -\frac{1}{5\p^3}(\pt\p)^4\Box\p \nn \\ && +\frac{1}{5\p^3}(\pt\p)^2\p^{,\mu}\p^{,\nu}\p_{,\mu\nu} -\frac{3}{20\p^4}(\pt\p)^6 \nn  \\ &=& -\frac{1}{4\p^2}\pt_\mu(\pt\p)^2\pt^\mu(\pt\p)^2 +\frac{1}{\p^2}\Box\p\p^{,\mu}\p^{,\nu}\p_{,\mu\nu} -\frac{1}{4\p^3}(\pt\p)^4\Box\p\,.
\label{67}
\eea
The last expression consists of only three terms and is particularly simple. We will focus on this version, and construct supersymmetric extensions for each of its three terms. For the first term, consider
\bea
{\cal L}_{4,\,{\rm 1st}\;{\rm term}}^{\rm SUSY}&=& \frac{1}{64(\P+\Pd)^2} \{ D,\Db \}(D\P D\P) \{ D,\Db \} (\Db\Pd\Db\Pd) \Big\vert_{\th\th\tb\tb}  \ .
 \label{68}
\eea
In components, up to quadratic order in fields other than $\p$ and using integration by parts, this becomes
\bea
{\cal L}_{4,\,{\rm 1st}\;{\rm term},\, {\rm quad}}^{\rm SUSY} &=& -\frac{1}{4\p^2}\pt_\mu (\pt\p)^2 \pt^\mu (\pt\p)^2  \nn \\&& + \frac{1}{\p^3}\p^{,\mu} \pt_\mu (\pt\p)^2 (\pt\chi)^2-\frac{1}{\p^2}\Box (\pt\p)^2 (\pt\chi)^2 -\frac{1}{\p^2} \chi_{,\mu} \chi_{,\nu} {\p_{,\l}}^\mu \p^{,\nu\l} \nn \\
&& +\frac{1}{\p^2}(\pt\p)^2\pt F \cdot \pt \Fs + \frac{1}{2\p^2}\pt_\mu (\pt\p)^2 \pt^\mu (\Fs F) + \frac{1}{\p^2}\p^{,\m\n}\p_{,\m\n}\Fs F \nn \\ &&
+ \frac{i}{2\p^3}\pt_\m (\pt\p)^2 \left( \pt^\m(\psi \s^\n \p_{,\n})\pb - \psi \s^\n \pt^\m (\pb \p_{,\n}) \right)\nn \\ &&+ \frac{i}{4\p^2}\pt_\m(\p_{,\n}\psi)\s^\n \sb^\l \s^\k \pt^\m(\pb_{,\k}\p_{,\l}) - \frac{i}{4\p^2}\pt_\m(\p_{,\n}\psi_{,\l})\s^\l \sb^\n \s^\k \pt^\m(\pb \p_{,\k}) \nn \\ &&-\frac{i}{2\p^2}\pt_\m(\p_{,\n}\psi)\s^\n \pt^\m(\pb_{,\l}\p^{,\l})+\frac{i}{2\p^2}\pt_\m(\p_{,\n}\psi^{,\n})\s^\l \pt^\m(\pb \p_{,\l})\nn \\ && + \frac{i}{4\p^2}\pt_\m(\p_{,\n}\psi)\s^\n \pt^\m(\pb \Box \p) -\frac{i}{4\p^2} \pt_\m(\Box \p \psi) \s^\n \pt^\m (\pb \p_{,\n}) \,,
 \label{69}
\eea
which reduces to the first term in (\ref{67}) when $\chi=\psi=F=0$.
Moreover, on a ghost condensate background, and dropping higher-derivative kinetic terms for fields other than $\p$,~(\ref{69}) further simplifies to
\bea
{\cal L}_{4,\,{\rm 1st}\;{\rm term},\, {\rm quad},\, X={\rm const}}^{\rm SUSY}  &=& -\frac{1}{4\p^2}\pt_\mu (\pt\p)^2 \pt^\mu (\pt\p)^2 +\frac{1}{\p^2}(\pt\p)^2\pt F \cdot \pt \Fs \nn \\ &&-\frac{9i}{4\p^4}(\pt\p)^2 \p_{,\m}\p_{,\n}(\psi^{,\m}\s^\n \pb - \psi \s^\n \bar\psi^{,\m}).
\label{70}
\eea
There are three noteworthy features here. First, we see that the scalar $\chi$ does not contribute at quadratic order on a constant $X$ background. Second, note that (\ref{70}) contains a kinetic term for the ``auxiliary'' field $F.$ And third, on a purely time-dependent background, the fermion kinetic term becomes non-covariant since it contains only the time-derivative part. These last two issues will be discussed in detail below.

The second term in the last line in (\ref{67}) can be supersymmetrized as
\bea
\nn
{\cal L}_{4,\,{\rm 2nd}\;{\rm term}}^{\rm SUSY}  &=& \frac{-1}{128(\P +\Pd)^2}\left(\{ D,\Db \}(\P + \Pd) \{ D,\Db \}( D\P D\P) \Db^2 \Pd +{\rm h.c.}\right) \bigg\vert_{\th\th\tb\tb}\,. \\
\label{71}
\eea
In terms of $\p$ and $\chi$, and using integration by parts, we obtain to quadratic order in fields other than $\p$,
\bea
{\cal L}_{4,\,{\rm 2nd}\;{\rm term},\, {\rm quad}}^{\rm SUSY}  &=&\frac{1}{2\p^2}\p^{,\mu}\pt_\mu (\pt\p)^2\Box\p +\frac{1}{2\p^2}(\pt\chi)^2\left[-\frac{2}{\p}(\pt\p)^2\Box\p+(\Box\p)^2+\p^{,\mu}\pt_\mu\Box\p\right] \nn \\
&+& \frac{1}{2\p^2}\pt_\mu(\pt\p)^2\chi^{,\mu}\Box\chi + \frac{1}{\p^2}\p^{,\mu}\p^{,\nu}\chi_{,\mu\nu}\Box\chi \nn \\ &+&  \frac{1}{2\p^2}\p^{,\m} \bigg[-(F\Box \p)_{,\m} \Fs - (\Fs \Box \p)_{,\m} F + (\pt F \cdot \pt \p)_{,\m}\Fs \nn \\ && \qquad \qquad + (\pt \Fs \cdot \pt\p)_{,\m} F - (F \p_{,\n})_{,\m}F^{\ast,\n} - (\Fs \p_{,\n})_{,\m} F^{,\n}\bigg] \nn \\ &+&  \frac{1}{4\p^2}\pt_\m (\pt\p)^2 \pt^\m (\Fs F ) - \frac{1}{\p^3} \p^{,\m}\pt_\m (\pt\p)^2 \Fs F \nn \\ & +&\frac{i}{4\p^2}\pt_\m \p \left( \pt^\m(\p_{,\n}\psi_{,\l})\s^\l\sb^\n\s^\k \pb_{,\k} - \psi_{,\k}\s^\k \sb^\n \s^\l \pt^\m(\pb_{,\l}\p_{,\n})\right) \nn \\ & +&\frac{i}{2\p^2}\pt_\m \p \left( \psi_{,\l}\s^\l \pt^\m(\bar\psi^{,\n}\p_{,\n}) - \pt^\m(\p^{,\n}\psi_{,\n})\s^\l \pb_{,\l}\right) \nn \\ &+&\frac{i}{4\p^2} \pt_\m \p \left( \pt^\m(\Box \p \psi) \s^\n \pb_{,\n} - \psi_{,\n}\s^\n \pt^\m(\pb \Box \p) + \pt^\m (\p_{,\n}\psi)\s^\n \Box \pb - \Box \psi \s^\n \pt^\m(\pb \p_{,\n})\right) \nn \\ &+&\frac{i}{4\p^2} \Box \p \left( \pt_\m(\p_{,\n}\psi)\s^\n \bar\psi^{,\m} -\psi^{,\m}\s^\n \pt_\m(\pb \p_{,\n}) \right) \nn \\ &+&\frac{i}{4\p^2}\pt^\m (\pt\p)^2 \left( \psi_{,\n}\s^\n \pb_{,\m} - \psi_{,\m}\s^\n \pb_{,\n}\right) \nn \\ &+& \frac{i}{2\p^3}\Box \p \pt_\m \p \left( \psi \s^\n \pt^\m(\pb \p_{,\n}) - \pt^\m(\p_{,\n}\psi)\s^\n \pb \right) \nn \\ &+& \frac{i}{2\p^3}\pt_\m \p \pt^\m (\pt\p)^2 \left( \psi \s^\n \pb_{,\n} - \psi_{,\n}\s^\n \pb\right) \,.
\label{72}
\eea
When $\chi=\psi=F=0$, this is simply the second term in the final line of (\ref{67}).
On a ghost condensate background, dropping higher-derivative kinetic terms for $\psi$ and integrating by parts, this further reduces to
 \bea
 {\cal L}_{4,\,{\rm 2nd}\;{\rm term},\, {\rm quad},\, X={\rm const}}^{\rm SUSY}  &=& \frac{1}{2\p^2}\p^{,\mu}\pt_\mu (\pt\p)^2\Box\p + \frac{3}{\p^3}(\pt\p)^4 \Fs F - \frac{2}{\p^2}(\pt\p \cdot \pt F)(\pt\p\cdot\pt\Fs) \nn \\ && + \frac{9i}{4\p^4}(\pt\p)^4( \psi_{,\n}\s^\n \pb - \psi \s^\n \pb_{,\n})\nn \\ && +\frac{3i}{4\p^4}(\pt\p)^2 \p_{,\m}\p_{,\n}(\psi^{,\m}\s^\n \pb - \psi \s^\n \bar\psi^{,\m}) \,.
 \label{73}
\eea
The expression above contains a non-Lorentz-covariant kinetic term for $F$ on a time-dependent background, as well as both a covariant and a non-covariant kinetic term for $\psi$.

For the third term in the last line of (\ref{67}), we have a choice of how to supersymmetrize since one can take either $(\pt\p)^4$ or $(\pt\p)^2\Box\p$ as the basic building block. This leads to {\it inequivalent} results.

\noindent {\bf Choice 1} --- based on $(\pt\p)^4$: The supersymmetric extension in this case is given by
\bea
{\cal L}_{4,\,{\rm 3rd}\;{\rm term}({\bf 1}),\, {\rm quad}}^{\rm SUSY}
&=& \frac{1}{64(\P+\Pd)^3} D\P D\P\Db\Pd\Db\Pd\{ D,\Db \} \{ D,\Db \}(\P + \Pd)  \Big\vert_{\th\th\tb\tb,{\rm quad}}  \nn\\ &=& -\frac{1}{4\p^3}\Box\p\left[(\pt \p)^4 + (\pt \chi)^4 - 2 (\pt \p)^2(\pt \chi)^2 + 4(\pt \p\cdot\pt\chi)^2 - (\pt\p)^2 \Fs F\right] \nn \\ && + \frac{i}{4\p^3}\Box \p [(\pt\p)^2(\psi_{,\n}\s^\n \pb - \psi \s^\n \pb_{,\n}) +2 \p_{,\m}\p_{,\n}(\psi^{,\m}\s^\n \pb - \psi \s^\n \bar\psi^{,\n})] \nn \\ && + \frac{i}{4\p^3}(\pt\p)^2\pt_\m \p (\psi \s^\m \Box \pb - \Box \psi \s^\m \pb) \,.
\label{74}
\eea
For $\chi=\psi=F=0$, this gives the third term in the final line of (\ref{67}).
Because of the $\Box\p$ prefactor, quadratic fluctuations in $\chi$ and $F$ vanish on a ghost condensate background and, dropping higher-derivative terms for $\psi$, we are left with
 \bea
 {\cal L}_{4,\,{\rm 3rd}\;{\rm term}({\bf 1}),\, {\rm quad},\, X={\rm const}}^{\rm SUSY} &=& -\frac{1}{4\p^3}(\pt\p)^4\Box\p \nn \\ && -\frac{3i}{4\p^4}(\pt\p)^2 \p_{,\m}\p_{,\n}(\psi^{,\m}\s^\n \pb - \psi \s^\n \bar\psi^{,\m})\,.
 \label{75}
 \eea

\noindent {\bf Choice 2} --- based on $(\pt\p)^2\Box\p$: Here we have additional freedom since there is some ambiguity about how to write the supersymmetric version of the $(\pt\p)^2$ factor. That is,
\bea
{\cal L}_{4,\,{\rm 3rd}\;{\rm term}({\bf 2}),\, {\rm quad}}^{\rm SUSY} &=& \frac{( D\P D\P \Db^2 \Pd + {\rm h.c.})}{64(\P + \Pd)^3} \bigg (\frac{(1-a)}{4}\{ D,\Db \}(\P + \Pd) \{ D,\Db \}(\P + \Pd) \nn \\
&& \qquad \qquad \qquad \qquad \qquad +a\{ D,\Db \} \P \{ D,\Db \} \Pd\bigg)\bigg\vert_{\th\th\tb\tb,{\rm quad}} \nn \\
&=&  -\frac{1}{4\p^3}\left(\Box \p (\pt\p)^2 -\Box \p (\pt\chi)^2 + 2\Box \chi \pt\p\cdot\pt\chi -2\Box\p \Fs F\right)\nn \\
&& \times \left((\pt\p)^2 + a (\pt\chi)^2\right) \nn \\ && + \frac{i}{8\p^3}(\pt\p)^2 \pt_\m \p [4(\psi^{,\m}\s^\n \pb_{,\n}-\psi_{,\n}\s^\n \bar\psi^{,\m})  + \Box \psi \s^\m \pb - \psi \s^\m \Box \pb] \nn \\ && + \frac{i}{8\p^3}(\pt\p)^2 \Box \p (\psi_{,\n}\s^\n \pb - \psi \s^\n \pb_{,\n})  + \frac{3i}{8\p^4}(\pt\p)^4(\psi_{,\n}\s^\n \pb - \psi \s^\n \pb_{,\n}) \nn \\ && + \frac{i}{4\p^3}\Box \p \p_{,\m}\p_{,\n}(\psi^{,\m}\s^\n \pb - \psi \s^\n \bar\psi^{,\m})\,,
\label{76}
\eea
where $a$ is an arbitrary real number. Note that for $\chi=\psi=F=0$, this also leads to the third term in the last line of (\ref{67}). Specializing to a ghost condensate background, integrating by parts and dropping higher-derivative kinetic terms for $\psi$, (\ref{76}) becomes
\bea
 {\cal L}_{4,\,{\rm 3rd}\;{\rm term}({\bf 2}),\, {\rm quad},\, X={\rm const}}^{\rm SUSY} &=& -\frac{1}{4\p^3}(\pt\p)^4\Box\p -\frac{3}{2\p^4}(\pt\p)^2(\pt\p\cdot\pt\chi)^2 + \frac{3}{4\p^4} (\pt\p)^4(\pt\chi)^2 \nn \\ && - \frac{3i}{8\p^4}(\pt\p)^4( \psi_{,\n}\s^\n \pb - \psi \s^\n \pb_{,\n}) \nn  \\ && +\frac{9i}{8\p^4}(\pt\p)^2 \p_{,\m}\p_{,\n}(\psi^{,\m}\s^\n \pb - \psi \s^\n \bar\psi^{,\m}) \,.
\label{77}
\eea
The terms proportional to $a$ have disappeared and, hence, one can choose a convenient value for $a$, such as $a=1$. Note that the term containing $F^{*}F$ is also missing. The fermion appears both with a canonical and a Lorentz-breaking fluctuation term on a constant $X$ background. Note that, contrary to all previous terms, ${\cal L}_{4,\,{\rm 3rd}\;{\rm term}({\bf 2})}^{\rm SUSY}$ introduces kinetic terms for the second scalar $\chi,$ both covariant and non-covariant.

In general, one can use a linear combination of Choices 1 and 2 above. We therefore define our supersymmetric generalization of ${\cal L}_{4}$ to be
\begin{equation}
{\cal L}_{4}^{\rm SUSY}={\cal L}_{4,\,{\rm 1st}\;{\rm term}}^{\rm SUSY}+{\cal L}_{4,\,{\rm 2nd}\;{\rm term}}^{\rm SUSY}+(1-b){\cal L}_{4,\,{\rm 3rd}\;{\rm term}({\bf 1})}^{\rm SUSY} + b \,{\cal L}_{4,\,{\rm 3rd}\;{\rm term}({\bf 2})}^{\rm SUSY}\, ,
\label{78}
\end{equation}
where $b$ is an arbitrary real number. In components, and restricted to a ghost condensate background with quadratic fluctuation terms up to two-derivatives,
this becomes
\bea
{\cal L}_{4,{\rm quad},X = {\rm const}}^{\rm SUSY} &=& {\cal L}_4 (\p) -\frac{3b}{2\p^4}(\pt\p)^2(\pt\p\cdot\pt\chi)^2 + \frac{3b}{4\p^4} (\pt\p)^4(\pt\chi)^2 \nn \\ && +\frac{3}{\p^3}(\pt\p)^4 \Fs F + \frac{1}{\p^2}(\pt\p)^2 \pt F \cdot \pt \Fs - \frac{2}{\p^2}(\pt\p \cdot \pt F)(\pt\p \cdot \pt \Fs) \nn \\ && + \left(\frac{18-3b}{8}\right)\frac{i}{\p^4}(\pt\p)^4( \psi_{,\n}\s^\n \pb - \psi \s^\n \pb_{,\n}) \nn  \\ && + \left(\frac{-18+15b}{8}\right)\frac{i}{\p^4}(\pt\p)^2 \p_{,\m}\p_{,\n}(\psi^{,\m}\s^\n \pb - \psi \s^\n \bar\psi^{,\m}) \,.
\label{78a}
\eea
When $\chi=\psi=F=0$ this reduces to the $\phi$-dependent Galileon term ${\cal L}_{4}$ in (\ref{67}), as it should. However, as it stands, the above Lagrangian still contains a number of troubling features. We now discuss how to address these.

First, the scalar $\chi$ appears with a non-covariant kinetic term. This, however, is not a serious problem  since it can be dealt with in exactly the same manner as discussed in and below (\ref{22}). That is, the non-covariant term can be eliminated by adding
\bea
&& - \frac{3b}{2^9(\P + \Pd)^4}D \P D \P \Db \Pd \Db \Pd \left(\{D,\Db\}(\P + \Pd)\{D,\Db\}(\P - \Pd)\right)^2 \bigg\vert_{\th\th\tb\tb, \, {\rm quad}} \nn \\
&&= -\frac{3b}{2\p^4}(\pt\p)^4(\pt\p\cdot\pt\chi)^2 \, .
\eea
Moreover, the coefficient in front of the covariant kinetic term for $\chi$ can be modified arbitrarily by adding a term proportional to
\bea
&& \frac{1}{2^8(\P + \Pd)^4}D \P D \P \Db \Pd \Db \Pd \left(\{D,\Db\}(\P - \Pd)\{D,\Db\}(\Pd - \P)\right) \nn
\\ &&=-\frac{1}{\p^4}(\pt\p)^4(\pt\chi)^2 \,
\eea
without affecting anything else. Hence, the stability of $\chi$ can always be assured.

Second, there are two troubling terms for the ``auxiliary'' field $F$, namely
\be
\frac{1}{\p^2}(\pt\p)^2 \pt F \cdot \pt \Fs - \frac{2}{\p^2}(\pt\p \cdot \pt F)(\pt\p \cdot \pt \Fs) \ .
\label{79}
\ee
These act as {\it kinetic terms for $F$} which, hence, is no longer an auxiliary field. We note that this occurs because we are interested in a time-dependent background $\p(t)$. Around the usual zero vacuum, such terms would be higher-order interactions and would not trouble us unduly. A propagating $F$-field is not {\it necessarily} a problem. It would imply that {\it both} complex components of the Weyl spinor $\psi$ would now propagate --- giving a supersymmetric  multiplet with four bosonic  and four fermionic physical degrees of freedom. However, in this paper we will follow a conservative approach and add the appropriate terms that restore $F$ to its auxiliary field status.
Consider the following supersymmetric terms evaluated for constant $X$,
\bea
&& -\frac{1}{2^8(\P + \Pd)^2}D\P D\P \Db \Pd \Db \Pd \{D,\Db\}D^2 \P \{D,\Db\}\Db^2 \Pd \Big\vert_{\th\th\tb\tb,{\rm quad}}\nn \\
& &\qquad = \frac{8}{(A + \As)^2}(\pt A)^2 (\pt \As)^2 \pt F \cdot \pt \Fs \nn \\ & & \qquad = \frac{1}{\p^2}(\pt \p)^4 \pt F \cdot \pt \Fs \,,
\label{80}
\eea
and
\bea
&& \frac{1}{2^{10}(\P + \Pd)^2} D \P D \P \Db \Pd \Db \Pd \vert \{ D,\Db \} \P \{ D, \Db \} D^2 \P \vert^2 \Big\vert_{\th\th\tb\tb,{\rm quad}} \nn \\
&&\qquad =  \frac{16}{(A + \As)^2} (\pt A)^2 (\pt \As)^2 (\pt A \cdot \pt F)(\pt \As \cdot \pt \Fs)\nn \\ && \qquad = \frac{1}{\p^2}(\pt \p)^4 (\pt\p\cdot\pt F)(\pt\p\cdot\pt\Fs)\,.
 \label{81}
\eea
At quadratic order, these do not involve $\chi$ or $\psi$.  Therefore, they can be added with suitable coefficients to ${\cal L}_{4}^{\rm SUSY}$ to {\it cancel} the unwanted kinetic terms for $F$, again without changing anything else. Thus, one can ensure that the auxiliary field remains truly auxiliary.

Finally, consider the fermionic kinetic terms in (\ref{78a}). The first is covariant, and unproblematic. The second one is Lorentz-violating and, hence, undesirable. This term can be eliminated by choosing  $b=6/5.$ With this choice, and adding in the terms just discussed, we find that a healthy supersymmetric extension of the fourth-order conformal Galileon Lagrangian is given by
\bea
\hat{{\cal L}}_4^{\rm SUSY} &=& \bigg(\frac{1}{64(\P+\Pd)^2} \{ D,\Db \}(D\P D\P) \{ D,\Db \} (\Db\Pd\Db\Pd) \nn \\
&-& \frac{1}{128(\P +\Pd)^2}\left[\{ D,\Db \}(\P + \Pd) \{ D,\Db \}( D\P D\P) \Db^2 \Pd +{\rm h.c.}\right] \nn \\ &-& \frac{1}{5\times 64(\P+\Pd)^3} D\P D\P\Db\Pd\Db\Pd\{ D,\Db \} \{ D,\Db \}(\P + \Pd) \nn \\ &+& \frac{6}{5\times 64(\P + \Pd)^3}( D\P D\P \Db^2 \Pd + {\rm h.c.}) \{ D,\Db \} \P \{ D,\Db \} \Pd \nn \\ &-& \frac{9}{2^8 \times 5(\P + \Pd)^4}D \P D \P \Db \Pd \Db \Pd \left(\{D,\Db\}(\P + \Pd)\{D,\Db\}(\P - \Pd)\right)^2 \nn \\ &+& \frac{1}{2^8(\P + \Pd)^2}D\P D\P \Db \Pd \Db \Pd \{D,\Db\}D^2 \P \{D,\Db\}\Db^2 \Pd \nn \\ &-& \frac{1}{2^{9}(\P + \Pd)^2} D \P D \P \Db \Pd \Db \Pd \left\vert \{ D,\Db \} \P \{ D, \Db \} D^2 \P \right\vert^2\bigg)\bigg\vert_{\th\th\tb\tb} \ .
\label{82}
\eea
In components, up to quadratic order in fields other than $\p$ on a constant $X$ background, this reduces to
\bea
\hat{{\cal L}}_{4,\, {\rm quad},\,X={\rm const}}^{\rm SUSY} &=& -\frac{1}{4\p^2}\pt_\mu(\pt\p)^2\pt^\mu(\pt\p)^2 +\frac{1}{\p^2}\Box\p\p^{,\mu}\p^{,\nu}\p_{,\mu\nu} - \frac{1}{4\p^3}(\pt\p)^4\Box\p\nn \\ &&  + \frac{9}{10\p^4} (\pt\p)^4(\pt\chi)^2 +\frac{3}{\p^3}(\pt\p)^4 \Fs F  + \frac{9i}{5\p^4}(\pt\p)^4( \psi_{,\n}\s^\n \pb - \psi \s^\n \pb_{,\n}) \,. \nn \\
 \label{83}
\eea
It is encouraging to see that healthy supersymmetric extensions of the Galileon Lagrangians exist, as demonstrated above. We would like to emphasize once more that, as should be clear already from the discussion around~(\ref{67}),
the supersymmetric extension of ${\cal L}_4$ given above is not unique. Hence it would be interesting to see, should a derivation of a supersymmetric Galileon theory be found in a more fundamental setting, precisely which form of the Lagrangian would arise.

\section{Discussion and Outlook} \label{discussion}

In this paper, we have shown how to construct supersymmetric extensions of the conformal Galileon theories. In doing so, we have uncovered a deep connection between Galileon and ghost condensate theories. That is, conformal Galileons can be seen as equivalent to ghost condensate models --- in terms of the temporal gradients alone, the two theories are {\it identical} up to an overall factor of $\phi^{-4}$ ---  but with improved behavior of the spatial gradients. This connection clarifies the role of both theories, and significantly simplifies the analysis of time-dependent solutions of the Galileon theories.

In our analysis, we have encountered two important subtleties, one related to supersymmetry and the other inherent already in the bosonic Galileons.
For the quadratic and cubic Galileon Lagrangians, the supersymmetric extensions are highly constrained and, around a ghost-condensate/self-accelerating-de-Sitter background, lead to covariant fluctuations for all fields. In contrast, for the quartic (and quintic) conformal Galileon Lagrangian there are many inequivalent ways to construct supersymmetric extensions. For some of these options, non-covariant fluctuations in some fields can arise, as well as kinetic terms for the ``auxiliary'' field. We have discussed these possibilities, and have provided an illustrative example of a completely healthy supersymmetric version of the fourth-order conformal Galileons, for which all fluctuations are covariant, and where the auxiliary field remains truly auxiliary.

A second subtlety we encountered, and which is inherent in higher-derivative theories, is that Lagrangians related using integration by parts generically lead to different stress-energy tensors and thus, for example, different conditions for violating the NEC. Keeping this subtlety in mind, let us now put all our results together and see if we can truly have a stable, NEC-violating solution of our supersymmetric conformal Galileon theory. For the Lagrangian
\be
{\cal L}^{\rm SUSY} = c_2 {\cal L}_2^{\rm SUSY} + c_3 {\cal L}_3^{\rm SUSY} + c_4 {\cal L}_4^{\rm SUSY},
\ee
with
${\cal L}_2^{\rm SUSY}$ given by (\ref{61}), ${\cal L}_3^{\rm SUSY}$ by (\ref{64}) and ${\cal L}_4^{\rm SUSY}$ by (\ref{82}), the conditions for having a) a ghost-condensate/self-accelerating-de-Sitter solution, b) stability, c)  NEC-violation and d) canonical and correct-sign fermionic fluctuations are
\bea
c_2 -\frac{3}{2} c_3 H_0^2 + \frac{3}{2}c_4 H_0^4 &=& 0 \label{eom}\\ c_2 - 3 c_3 H_0^2 + \frac{9}{2} c_4 H_0^4 &>& 0 \\ c_2 + \frac{3}{2} c_3 H_0^2 + \frac{3}{2}c_4 H_0^4 &<& 0 \\
c_2 - \frac{3}{2}c_3H_0^2 + \frac{18}{5} c_4 H_0^4 &>& 0 \ ,
\eea
respectively. These can be satisfied simultaneously provided that
\be
c_2 < \frac{3}{2} c_3 H_0^2 < 0,
\ee
with the value of $c_4 H_0^4$ determined by (\ref{eom}). Hence, we have an example of a supersymmetric conformal Galileon theory that has a background solution which is both stable and can violate the NEC at the same time! We note that this is merely a proof of principle, and that
using a supersymmetric ${\cal L}_5,$ or other choices for supersymmetrizing ${\cal L}_4,$ will lead to a variety of such theories. For these, the conditions for violating the NEC and having canonical fermionic fluctuations will have to be re-evaluated on a case-by-case basis, but it seems likely to us that healthy supersymmetric models might also exist for which all the conditions mentioned above can be satisfied with $c_2>0$. In that case, even the ordinary zero vacuum would be stable.

We anticipate a number of applications for our results:

\begin{itemize}

\item Since the (non-supersymmetric) conformal Galileon theories can be derived as the Lagrangians describing the fluctuations of a brane in a higher-dimensional space-time, there does not seem to exist a fundamental obstacle to deriving the supersymmetric Galileons in a supergravity context. It will then be interesting to see precisely which version of the supersymmetric Galileons comes out naturally. Moreover, such a treatment would require one to extend the present work to local supersymmetry and the coupling to gravity, which will be of importance for cosmological applications. This derivation will appear elsewhere~\cite{toappear2}.

\item As we discussed in detail in the paper, two Galileon actions, related using integration by parts and dropping the surface terms, are physically inequivalent in that they lead to the same time-dependent backgrounds, but to different pressures. Hence, for two such theories, the conditions for violating the NEC are different. It may be that one theory allows for stable violations of the NEC, while the other does not. It will be interesting to investigate this situation further in a cosmological context. Indeed, it means that in approaching a regime where the NEC is violated, such as a cosmic bounce, spatial gradients must necessarily make their presence felt, and either allow or disallow entering into the NEC-violating regime. This should be the case regardless of how small the spatial gradients are initially! It will be of interest to see how this works out in a concrete model.

\end{itemize}

Supersymmetric Galileons provide a fascinating theoretical laboratory in which to study the connections between higher derivatives, supersymmetry and violations of the NEC. This is at present largely unchartered territory, but, in this paper, we hope to have provided the basic tools necessary for mapping it out.

\section*{Acknowledgments}

We would like to thank Kurt Hinterbichler and Alberto Nicolis for useful discussions. J.L.L. expresses his thanks to the University of Pennsylvania, for its hospitality while this work was being completed. J.K. and B.A.O. are supported in part by the DOE under contract No. DE-AC02-76-ER-03071, the NSF under grant No. 1001296, and by the Alfred P. Sloan Foundation (JK). J.L.L. is supported by a Starting Grant from the European Research Council.

\section*{Appendix A: Useful Superfield Expressions}

In constructing the supersymmetric extensions of the higher-derivative scalar actions discussed in this paper, we make frequent use of a number of building blocks. We list them here for reference, our notation and conventions being those of Wess and Bagger \cite{Wess:1992cp}:
\bea
D\P D \P &=& 2 \psi\psi + 4 \sqrt{2} F \th\psi + 4 \sqrt{2} i \psi \s^\m \tb A_{,\m} + 4 \th\th F^2 -4 \tb\tb (\pt A)^2 \nn \\ && + 8i\th\s^\m\tb F A_{,\m}-4i \th\psi \psi_{,\m}\s^\m \tb + 4i \th \psi_{,\m} \psi \s^\m \tb \nn \\ && + 2 \sqrt{2} \tb\tb(\th\psi \Box A + \psi_{,\m}\s^\m \sb^\n \th A_{,\n} -2A^{,\m}\th\psi_{,\m}) \nn \\ && +2\sqrt{2} i \th\th(F_{,\m}\psi\s^\m \tb - 3 F \psi_{,\m} \s^\m \tb) \nn \\ && + \th\th\tb\tb (4F\Box A - 4 \pt F \cdot \pt A - \psi \Box \psi + \pt \psi \cdot \pt \psi - \psi_{,\m} \s^\m \sb^\n \psi_{,\n})\,,
\label{A1}
\\
D^2 \P &=& -4F + 4 \sqrt{2} i \psi_{,\m}\s^\m \tb -4 \tb\tb \Box A \nn \\ &&+ 4i \th\s^\m \tb F_{,\m} -2\sqrt{2} \tb\tb \th \Box \psi -\th\th\tb\tb \Box F\,.
\label{A2}
\eea
Multiplying the first of these building blocks with the hermitian conjugate of the second, we obtain, up to quadratic order in the spinor $\psi$,
\bea
D\P D \P \Db^2 \Pd &=& -8 \psi\psi \Fs -16 \sqrt{2} \th\psi F \Fs -16\sqrt{2} i \psi \s^\m \tb A_{,\m} \Fs \nn \\ && -8 \th\th \psi\psi \Box \As -16 \th\th F^2 \Fs + 16i \th\th \psi \s^\m \bar{\psi}_{,\m}F + 16 \tb\tb (\pt A)^2 \Fs \nn \\ && -32 i \th\s^\m \tb A_{,\m}F \Fs + 32 \psi\s^\m \tb \th \s^\n \bar\psi_{,\n}A_{,\m} \nn \\ && -8i \th\s^\m \tb \psi\psi F^\ast_{,\m} + 16i \th\psi \psi_{,\m}\s^\m \tb \Fs -16i \th \psi_{,\m}\psi \s^\m \tb \Fs \nn \\ && + 8\sqrt{2} \th\th \Big(-2i\psi\s^\m\tb A_{,\m}\Box \As -2 \s^\m \bar\psi_{,\m}\s^\n \tb F A_{,\n} \nn \\ && \qquad \qquad +  i\psi\s^\m\tb F F^\ast_{,\m}-i\psi\s^\m\tb F_{,\m}\Fs + 3i \psi_{,\m}\s^\m\tb F \Fs\Big) \nn \\ && + 8 \sqrt{2} \tb\tb \Big(2i \th \s^\m \bar\psi_{,\m}(\pt A)^2 + \th \s^\m \sb^\n \psi A_{,\n} F^\ast_{,\m} \nn \\ && \qquad \qquad +2 A^{,\m}\th\psi_{,\m}\Fs -\th\psi \Box A \Fs -\psi_{,\m}\s^\m \sb^\n \th A_{,\n}\Fs\Big) \nn \\ && + \th\th\tb\tb \Big(16 (\pt A)^2 \Box \As -16 \Box A F \Fs + 16 \Fs \pt F \cdot \pt A -16 F \pt \Fs \pt A \nn \\ && \qquad \quad + 8i \psi_{,\m}\s^\m \sb^\n \s^\l \bar\psi_{,\l} A_{,\n} -16i \psi_{,\m}\s^\n \bar\psi_{,\n}A^{,\m} + 8i \psi \s^\n \bar\psi_{,\n}\Box A \nn \\ && \qquad \quad + 8i \psi \s^\m \Box \bar\psi A_{,\m} + 4 \psi \Box \psi \Fs -4 \pt \psi \cdot \pt \psi \Fs + 4 \psi_{,\m}\s^\m \sb^\n \psi_{,\n}\Fs \nn \\ && \qquad \quad + 4 \psi \s^\m \sb^\n \psi_{,\n} F^\ast_{,\m} -4 \psi_{,\m}\s^\n \sb^\m \psi F^\ast_{,\n} -2 \psi\psi \Box \Fs\Big)\ .
\label{A3}
\eea
We also make frequent use of
\bea
(\P + \Pd)^k &=& (A + \As)^k + k\sqrt{2} (A+\As)^{k-1}(\th\psi + \tb \bar\psi) + k \th\th (A+\As)^{k-1} F \nn \\ && + k \tb\tb (A+\As)^{k-1}\Fs + ki\th\s^\m\tb (A+\As)^{k-1}(A_{,\m}-A^\ast_{,\m})\,,
\eea
where we have dropped the top component as well as terms quadratic and higher in fields other than $\p.$

\section*{Appendix B: Non Lorentz-Covariant Fermion Kinetic Terms}

In Sec.~\ref{fermioncure}, we showed that the exact linear combination of (\ref{32}) + (\ref{30}) not only gives the conformal third-order scalar Galileon Lagrangian, but also results in a Lorentz-invariant fermion kinetic term. In this Appendix, we generalize this analysis by allowing for a more general linear combination. The fermion kinetic term now breaks Lorentz invariance, and the resulting {\it generalized} Galileon theory is only invariant under dilations but not special conformal transformations.

Instead of (\ref{35}), consider the more general expression
\bea
&& (\ref{32})+(1+\Delta) \times (\ref{30})= \frac{1}{12\phi^{4}}(\partial\phi)^4 +\frac{1}{18 \phi^{2}}(\partial_{\mu}\partial_{\nu} \phi)^{2}-\frac{1}{18 \phi^{2}}(\Box\phi)^{2}+ \frac{\Delta}{4\phi^{4}}(\partial\phi)^4 \nn \\
&& \qquad -\frac{i(1+\Delta)}{4\phi^{4}}(\partial\phi)^2(\psi_{,\mu}\sigma^\mu {\bar{\psi}} - \psi \sigma^\mu {\bar{\psi}}_{,\mu})-\frac{i\Delta}{2\phi^{4}}\phi_{,\mu}\phi_{,\nu}(\psi^{,\nu}\sigma^\mu {\bar{\psi}} - \psi \sigma^\mu {\bar{\psi}}^{,\nu})+\ldots \nn \\
\label{a1}
\eea
where $\Delta$ is a constant. It follows that $3 \times (\ref{35})$, the higher-derivative term entering expression (\ref{38}), is now replaced by
\be
{\cal{L}}_{\Delta} -\frac{3i(1+\Delta)}{4\phi^{4}}(\partial\phi)^2(\psi_{,\mu}\sigma^\mu {\bar{\psi}} - \psi \sigma^\mu {\bar{\psi}}_{,\mu}) -\frac{3i\Delta}{2\phi^{4}}\phi_{,\mu}\phi_{,\nu}(\psi^{,\nu}\sigma^\mu {\bar{\psi}} - \psi \sigma^\mu {\bar{\psi}}^{,\nu})+\ldots
\label{a2}
\ee
where
\begin{equation}
{\cal{L}}_{\Delta}\equiv -{\cal{L}}_{3}+\frac{3\Delta}{4\phi^{4}}(\partial\phi)^4\,,
\label{a3}
\end{equation}
and ${\cal{L}}_{3}$ is the third-order conformal Galileon Lagrangian in (\ref{45}).

For $\Delta=0$, the last term in (\ref{a2}) vanishes, and the fermion kinetic term is Lorentz-covariant with the correct sign on the ghost condensate background. Furthermore, ${\cal{L}}_{\Delta=0}=-{\cal{L}}_{3}$
is the standard conformal Galileon action discussed in Sec.~\ref{galP(X)}. As mentioned in Sec.~4, ${\cal{L}}_{3}$ is invariant, up to a total derivative, under the infinitesimal dilations and special conformal transformations given in~(\ref{42a}). We now show that in the more general case, when $\Delta\neq 0$, only the dilation symmetry survives.

To see this, we work in terms of $\pi$ and decompose ${\cal{L}}_{3}$ as
\begin{equation}
{\cal{L}}_{3}={\cal{L}}_{3A}+{\cal{L}}_{3B}\,,
\label{a4}
\end{equation}
with
\begin{equation}
{\cal{L}}_{3A}=-\frac{1}{2}(\pt\pi)^2\Box\pi\,; \qquad {\cal{L}}_{3B}=- \frac{1}{4}(\pt\pi)^4
\label{a5}
\end{equation}
being of order ${\cal{O}}(\pi^3)$ and ${\cal{O}}(\pi^4)$ respectively. In this notation, the generalized Lagrangian ${\cal{L}}_{\Delta}$ can be written as
\be
{\cal{L}}_{\Delta}= - {\cal{L}}_{3A} -(1+3\Delta){\cal{L}}_{3B} \ .
\label{a11}
\ee
First consider the dilation transformation $\delta_{c}\pi=c\left(1+x^{\mu}\pt_{\mu}\pi\right)$. Since each $\pi$ in~(\ref{a5})
is acted on by at least one derivative, the relevant variation is
\begin{equation}
\delta_{c} \pt_{\mu} \pi =c\pt_{\mu} (x^{\alpha} \pt_{\alpha}\pi) \ .
\label{a6}
\end{equation}
This preserves the order in $\pi$ and, hence, the variations of ${\cal{L}}_{3A}$ and ${\cal{L}}_{3B}$ cannot cancel against each other. Instead they must be separately invariant (up to a total derivative) under this transformation. It is straightforward to check that this is indeed the case. And since ${\cal{L}}_{\Delta}$ is just a linear combination of these two terms, it too is dilation invariant.

Now consider the infinitesimal special conformal transformation, given by the second equation in~(\ref{42a}). The relevant variation in this case,
\begin{equation}
\delta_{v} \pt_{\mu} \pi=v_{\mu}-\partial_{\mu} \left( \pt_{\alpha}\pi \left(\frac{1}{2} v^{\alpha} x^{2}-(v \cdot x)x^{\alpha}\right)\right)\,,
\label{a8}
\end{equation}
has both $0$th- and $1$st-order contributions in $\pi$. Thus, unlike the previous transformation, the variations of the two terms in~(\ref{a5}) {\it can} cancel against each other,
so that neither need be a total divergence. For example, consider
\begin{equation}
\delta_{v} {\cal{L}}_{3B}=-v^{\mu}(\pt_{\mu}\pi)(\pt \pi)^{2} +{\cal{O}}(\pi^{4}) \label{a9} \ .
\end{equation}
Since $\delta_{v} {\cal{L}}_{3A}$ is at most cubic in $\pi$, the ${\cal{O}}(\pi^{4})$ term in~(\ref{a9}) must be a total derivative, which is indeed the case.
However,  the first term need {\it not} a total derivative, and, in fact, it is not. This is most easily checked by defining an action $-\int {\rm d}^4x\, v^{\mu}(\pt_{\mu}\pi)(\pt \pi)^{2}$. If the integrand were a total derivative, then the variation of this action would vanish identically. Instead we find the non-zero result
\begin{equation}
-\frac{\delta}{\delta \pi}\int {\rm d}^{4}x \, v^{\mu}(\pt_{\mu}\pi)(\pt \pi)^{2} =2v^{\mu}(\pt_{\mu}\pi)
 \Box \pi + 4v^{\mu}(\pt^{\alpha}\pi)(\pt_{\mu}\pt_{\alpha}\pi) \neq 0 \ .
\label{a10}
\end{equation}
It follows that ${\cal{L}}_{3B}$ is not by itself invariant under~(\ref{a8}). The invariance of ${\cal L}_3$ relies on a cancellation between the first term in (\ref{a9}) and the ${\cal{O}}(\pi^{3})$ part of the variation of ${\cal{L}}_{3A}$  (the ${\cal{O}}(\pi^{2})$ term in $\delta_{v} {\cal{L}}_{3A}$ is a total derivative).

The immediate corollary is that ${\cal{L}}_{\Delta}$ is {\it not} invariant under the transformation~(\ref{a8}) for $\Delta \neq 0$. In other words,
when the fermion kinetic term is {\it not} Lorentz invariant on the condensate background, the purely $\pi$-dependent part of Lagangian is still invariant
under dilations~(\ref{a6}) but breaks special conformation transformations~(\ref{a8}). This generalized class of ``detuned" Galileon theories, and their supersymmetric extension discussed above,
admit a ghost condensate vacuum and are potentially interesting in their own right, such as for cosmological applications. We will explore their properties elsewhere~\cite{toappear}.

\end{document}